\documentclass[a4paper,12pt]{article}

\usepackage{latexsym}
\usepackage{amsmath,amssymb,bm,ascmac,color,calc}
\usepackage[mathscr]{eucal}
\usepackage{graphicx}
\usepackage{cite}
\usepackage{lineno}

\newcommand{\ka}{\kappa}
\newcommand{\kaB}{\bar{\kappa}}
\newcommand{\la}{\lambda}
\newcommand{\qq}{q\bar{q}}
\newcommand{\sub}{\text{sub}}
\newcommand{\LL}{\text{LL}}
\newcommand{\De}{\Delta}
\newcommand{\DeT}{\tilde{\Delta}}
\newcommand{\PiT}{\tilde{\Pi}}
\newcommand{\GammaT}{\tilde{\Gamma}}
\newcommand{\asB}{\bar{\alpha}_s}

\newcommand{\Ein}{\text{Ein}}
\newcommand{\wB}{\overline{w}}
\newcommand{\Lnew}{\Lambda_{\rm new}}
\newcommand{\Njets}{N_{\rm jets}}


\begin{document}
\begin{titlepage}

\vspace*{2cm}

\begin{center}
{\large\bf 
Quark jet rates and quark/gluon discrimination in multi-jet final states
}\\[15mm]

\renewcommand{\thefootnote}{\fnsymbol{footnote}}
Yasuhito Sakaki\\ \bigskip
\renewcommand{\thefootnote}{\arabic{footnote}}

{\em Department of Physics, Korea Advanced Institute of Science and Technology, 291 Daehak-ro, Yuseong-gu, Daejeon 34141, Republic of Korea}
\end{center}

\vspace{1cm}
\begin{abstract}
We estimate the number of quark jets in QCD multi-jet final states at hadron colliders. 
In the estimation, we develop the calculation of jet rates into that of quark jet rates. 
From the calculation, we estimate the improvement on the signal-to-background ratio for a signal semi-analytically by applying quark/gluon discrimination, where the signal predicts many quark jets.
We introduce a variable related to jet flavors in multi-jet final states and propose a data-driven method using the variable.  
As the same with the semi-analytical result, the improvements on the signal-to-background ratio using the variable in Monte-Carlo analysis are estimated.

\end{abstract}
\end{titlepage}

\hrule
\tableofcontents
\vskip .2in
\hrule
\vskip .4in

\setlength{\parskip}{10pt}%

\renewcommand\thefootnote{\roman{footnote}}
\section{Introduction}
So far we have not caught a clear sign of physics beyond the standard model at the LHC.  We should maximize the discoverability of new physics at the LHC by using information of final states more precisely. 
In conventional analyses, we categorize events by inclusive variables like the number of jets, the scalar sum of the transverse momentum of jets and so on, then find signal regions using exclusive variables like the transverse momentum of objects, the distance between objects etc. 
We are able to access more specific features of events using jet substructure, and the related studies have developed dramatically in this ten years \cite{Abdesselam:2010pt,Altheimer:2012mn,Altheimer:2013yza,Adams:2015hiv,Gras:2017jty,Larkoski:2017jix}.

In the studies, many methods and variables to identify the origin of a jet using the jet substructure information have been proposed \cite{Berger:2003iw, Butterworth:2008iy, Almeida:2008yp, Thaler:2010tr, Jankowiak:2011qa, Gallicchio:2011xq, Thaler:2011gf, Krohn:2012fg, Gallicchio:2012ez, Larkoski:2013eya, Larkoski:2014gra, Larkoski:2014zma, Bhattacherjee:2015psa, Moult:2016cvt, Davighi:2017hok} and been understood well based on the resummation with the assumption of soft-collinear factorization \cite{Li:2011hy, Feige:2012vc, Waalewijn:2012sv, Bolzoni:2012ii, Dasgupta:2013ihk, Larkoski:2013paa, Larkoski:2014uqa, Larkoski:2014tva, Larkoski:2014pca, Procura:2014cba, Dasgupta:2015yua, Larkoski:2015kga, Dasgupta:2015lxh, Pietrulewicz:2016nwo, Frye:2016aiz, Dasgupta:2016ktv, Salam:2016yht, Kang:2017mda, Liu:2017pbb, Liu:2018ktv, Kang:2018jwa}.
Applications of the jet substructure techniques to new physics searches at the LHC are also considered (see e.g., \cite{CMS:2013kfa, FerreiradeLima:2016gcz, Bhattacherjee:2016bpy, Chakraborty:2017mbz, Park:2017rfb}).
Especially, numerous studies related to identifications of boosted jets originated from the top quark, Higgs and $Z$/$W$ bosons have been appeared. 
Some prominent variables were measured experimentally and the performance of tagging techniques have been tested.
The tagging techniques are being used for new physics searches at the LHC \cite{Aad:2013gja, Aad:2015rpa, TheATLAScollaboration:2015ynv, ATLAS:2016wlr}. 
These boosted jets have a multi-prong structure inside the jet and we can distinguish these from QCD jets using the features, where the QCD jets mean quark- and gluon-initiated jets.
There is also progress in the studies for the separation between the quark jets and gluon jets.  
The performance of separation and the shapes of variables have been measured \cite{
Aad:2011gn, Aad:2011sc, ATLAS:2012am, Chatrchyan:2012mec, Khachatryan:2014hpa, Aad:2014gea, Aad:2015cua, Aad:2016oit}. 
In recent years, applications of machine learning techniques to improve the separation performance are focused on \cite{
deOliveira:2015xxd, Komiske:2016rsd, Kasieczka:2017nvn, Louppe:2017ipp, Cohen:2017exh, Komiske:2017ubm, Butter:2017cot, Metodiev:2017vrx, Cheng:2017rdo, Luo:2017ncs, Komiske:2018oaa, Macaluso:2018tck, Fraser:2018ieu}.

One of the differences between the boosted jet tagging technique and the quark/gluon discrimination is the size of the jet radius used in the analysis. 
The multi-prong structure of boosted jets are formed by decays stem from the electroweak interaction, and a large jet radius is basically required to catch most of the decay products. 
The QCD jets have 1-prong structures such that there is a core parton carrying on most of the energy of jet, and the core is dressed in soft-gluons radiated from itself. 
The main difference between quark jets and gluon jets stems from the difference of color factors for the gluon radiation.  Gluon jets emit more partons and wider radiations due to the difference. 
Neglecting the logarithmic scaling on the strong coupling and masses of the active quarks, the QCD radiation is the scale-invariant. That is, if one zooms in on a QCD jet, one will find a repeated self-similar pattern of jets within jets within jets, reminiscent of fractals. 
The difference exists even in a neighborhood of the jet core, therefore the quark/gluon discrimination works out even if the jet radius is small.

Due to the properties, the quark/gluon discrimination is maximally utilized in multi-jet final states. 
In the case that a signal has more $n$ quark jets compared to backgrounds, we naively expect the signal-to-background ratio increases $(\epsilon_q/\epsilon_g)^n$ times using the quark/gluon tagging, where $\epsilon_{q}$ and $\epsilon_{g}$ are the quark and gluon jet efficiencies and $\epsilon_q/\epsilon_g>1$ in the assumption such that quarks come from a signal.

Many new physics models form multi-jet final states. 
For example, heavy colored resonances in each model, like gluino or squarks in SUSY, could emit many partons via their cascade decays, and there are several studies for these at the LHC \cite{
Aad:2012hm, Chatrchyan:2012lia, CMS:2014exa, Aad:2016jxj, Khachatryan:2016xim, Sirunyan:2017cwe, Aaboud:2017faq}. 
Another example is searches of low-scale gravity which deals with the hierarchy problem concerned with the difference between the electroweak and Plack scale.  The models can predict microscopic black holes or highly excited string states at TeV scale.  The objects emit a large number of energetic particles which are mostly quark and gluons, and the phenomenon is constrained experimentally \cite{
Chatrchyan:2013xva, Aad:2014gka, Aad:2015zra, Aad:2015mzg, Sirunyan:2017anm}. 
Moreover, the multi-jet final state is a good probe for the higher dimensional operators which are caused by new color interactions at a high energy scale.
There are two type of dimension-6 pure QCD operators,
$g_s f_{abc}G^{\mu}_{a\nu} G^{\nu}_{b\lambda} G^{\lambda}_{c\mu}$ and 
$g_s^2 (\bar{q}\gamma_{\mu}T_a q) (\bar{q}\gamma^{\mu}T_a q)$, 
and especially the triple gluon field strength gets large enhancement at high energy and large jet multiplicity regions \cite{
Cho:1994yu, Krauss:2016ely}. 
The operator also predicts a characteristic quark/gluon jet fraction such that $G^3$-operator forms leading and sub-leading gluon jets although the leading and sub-leading jets tend to be valence quark jets in the standard model backgrounds.

We develop the calculation of jet rates into {\it quark jet rates}, and estimate quark/gluon jet fractions in the QCD multi-jet final states. 
Also, we consider a data-driven analysis for new physics searches in multi-jet final states.
In the analysis, we introduce a variable defined in events having jets greater than or equal to $n$,
\begin{align}
d = \sqrt{ \frac{1}{n} \sum_{i=1}^{n} Q_i^2 }, 
\end{align}
where $Q_i(>0)$ is assigned to $i$-th jet and it becomes larger when the jet substructure seems to be a quark jet. 
So, $d$ gets larger if events have more quark jets. 
In conventional analyses for multi-jet final states, we fit a distribution of an inclusive variable like the scalar sum of the transverse momenta $H_T$ in a control region and predict the number of background in a signal region using the fit functions. 
We also fit remaining rates of the number of events after imposing $d$-cut for each $H_T$ bins as will be shown in Fig.~\ref{fig:dcut} and show a big improvement from conventional analyses by using the new information.

This paper is organized as follows. 
In Section~\ref{sec:Quark-jet-rates}, we calculate quark jet rates at hadron colliders based on the generating functional method. 
In Section~\ref{sec:Numerical-results}, we estimate how many quark jets are contained in the QCD multi-jet background using the formulae derived in the last section.  
Improvements of the signal-to-background ratio for the analysis in multi-jet final states by using the quark/gluon discrimination are also estimated semi-analytically. 
In Section~\ref{sec:BSM}, the improvements using the variable $d$ are estimated in Monte-Carlo analysis.
In Section~\ref{Conclusions}, we summarize our results and reach out to a conclusion.

\section{Quark jet rates in multi-jet final states}
\label{sec:Quark-jet-rates}

We first estimate how many quark jets are contained in the QCD multi-jet background at hadron colliders.  
The estimation is useful to know the impact of quark/gluon discrimination for new physics searches and helps to understand the results of the analysis. 
Assuming infinite calculation resources, we can add any number of additional partons into parton showers using the matching schemes \cite{
Catani:2001cc, Krauss:2002up, Mangano:2001xp, Lonnblad:2001iq, Lonnblad:2011xx} in the simulation of multi-jet final states.  
However, we don't have such enormous calculation resources. 
So, we use the generating functional method \cite{Konishi:1979cb, Dokshitzer:1991wu, Ellis:1991qj, Gerwick:2012fw} based on DGLAP equations. All of leading logarithmic (LL) terms and a part of next-to-leading logarithmic (NLL) terms are taken into account in the calculation. 
Matrix element corrections for additional partons are absence in the calculations and those effects are examined in Appendix \ref{app:MEC}.

The definition of quark and gluon jets are typically given by using parton-level information\cite{Banfi:2006hf, Gallicchio:2011xc, Buckley:2015gua}.  
We also use the parton-level definition in this study.  Although the definition is unphysical since jets are observed at hadron-level, we assume the number of quark jets defined at parton-level is close the number at hadron-level.  
It should be noted that a well-defined definition of quark/gluon jets at hadron-level was proposed recently \cite{Komiske:2018vkc}. 
We can see a result in the reference that quark jets fraction defined at parton-level can be extracted using hadron-level information.  It has been demonstrated for the hardest jet in $Z$+jet process and for the hardest two jets in dijet process with a Monte-Carlo event generator.

\subsection{Generating functionals}
Conventionally, a generating functional for a final state parton $i$ is defined as,
\begin{align}
\Phi_i(u,p,t)
&=\sum_{n=1}^{\infty} u^n R_n^{(i, {\rm out})}(p,t), \label{eq:conventional}
\end{align}
where $p$ and $t$ are the transverse momentum and energy scale for the parton $i$.  
We call a parton whose transverse momentum is larger than $p_0$ and which is separated from other partons by $R$ or more in $\eta$-$\phi$ plane as a jet, where the non-global logarithmic effect \cite{Dasgupta:2001sh} is ignored.
The jet rate $R_n^{(i, {\rm out})}$ represents the probability that the parton $i$ forms a $n$ jets configuration by final state radiation \cite{
Ellis:1985vn, Berends:1989cf, Berends:1990ax, Gerwick:2011tm, Englert:2011cg}. 
We develop the definition, 
\begin{align}
\Phi_i(u,v,p,t) = \sum_{n=1}^{\infty} u^n \mathcal{R}_{n}^{(i, {\rm out})}(v,p,t), 
\end{align}
where the modified jet rates are given by a polynomial expression,
\begin{align}
\mathcal{R}_{n}^{(i, {\rm out})}(v,p,t) = \sum_{m=0}^{n} v^m R_{n,m}^{(i, {\rm out})}(p,t). 
\end{align}
We call $R_{n,m}^{(i, {\rm out})}$ as {\it quark jet rates} and it represents the probability that $i$ forms a $n$ jets configuration in which $m$ quark jets are contained.\footnote{
You may prefer the definition of functional such that 
$\hat{\Phi} (u_g,u_q)=\sum _{n_g} \sum _{n_q} u_g^{n_g} u_q^{n_q} \hat{R}_{n_g,n_q}$, where $n_q$ and $n_g$ are the number of quark and gluon jets, and $\hat{R}_{n_g,n_q}$ is the probability that an event has $n_q$ quark jets and $n_g$ gluon jets.  The two functionals are just related by $\Phi (u,v)=\hat{\Phi}(u,u v)|_{n_q=n-n_g, \, n_g=m}$.
} 
The number of jet $n$ starts from $1$ since the final state itself becomes a jet even if it doesn't emit any resolved emissions. 
The jet rate $R_n^{(i, {\rm out})}$ in Eq.~(\ref{eq:conventional}) is simply given as,
\begin{align}
R_n^{(i, {\rm out})}(p,t) = \mathcal{R}_{n}^{(i, {\rm out})}(1,p,t). 
\end{align}
We can acquire the quark jet rates by differentiating the functional as,
\begin{align}
R_{n,m}^{(i, {\rm out})}(p,t) = \left. \frac{1}{n! m!} \frac{\partial^{n}}{\partial u^n} \frac{\partial^{m}}{\partial v^m} \Phi_i(u,v,p,t) \right\vert_{u=v=0}.
\label{gen_diff}
\end{align}

Similarly, we introduce a generating functional for an initial state parton $i$, 
\begin{align}
\Psi_{i}(u,v,x,t)
= \sum_{n=0}^{\infty} u^n \mathcal{R}_{n}^{(i, {\rm in})}(v,x,t), 
\end{align}
where 
\begin{align}
\mathcal{R}_{n}^{(i, {\rm in})}(v,x,t)
= \sum_{m=0}^{n} v^m R_{n,m}^{(i, {\rm in})}(x,t). 
\end{align}
Quark jet rates for the initial state $R_{n,m}^{(i, {\rm in})}$ represents the probability that $i$ emits $n$ jets in which $m$ quark jets are contained.
The argument $x$ is the energy fraction for $i$, therefore, the parton carry on the energy $x p_{\rm beam}$, where $p_{\rm beam}$ is the hadron beam energy.  The number of jet $n$ starts from $0$ since the initial state doesn't generate any jet if it doesn't emit any resolved emissions.

A generating functional for a hard process is given by a product of functionals for initial and final states.  For example, a generating functional which has initial states $i_1, i_2$ and final states $f_1, f_2$ is given as $\bold\Phi = \Psi_{i_1}\Psi_{i_2}\Phi_{f_1}\Phi_{f_2}$, and we can derive the quark jet rates for the hard process by the differentiate in Eq.~(\ref{gen_diff}).

For brevity, we omit the arguments $u$ and $v$ in the generating functionals below.

\subsubsection{Evolution equations}

\begin{figure}[!t]
\begin{center}
\includegraphics[width=0.5\linewidth, bb=0 0 580 200]{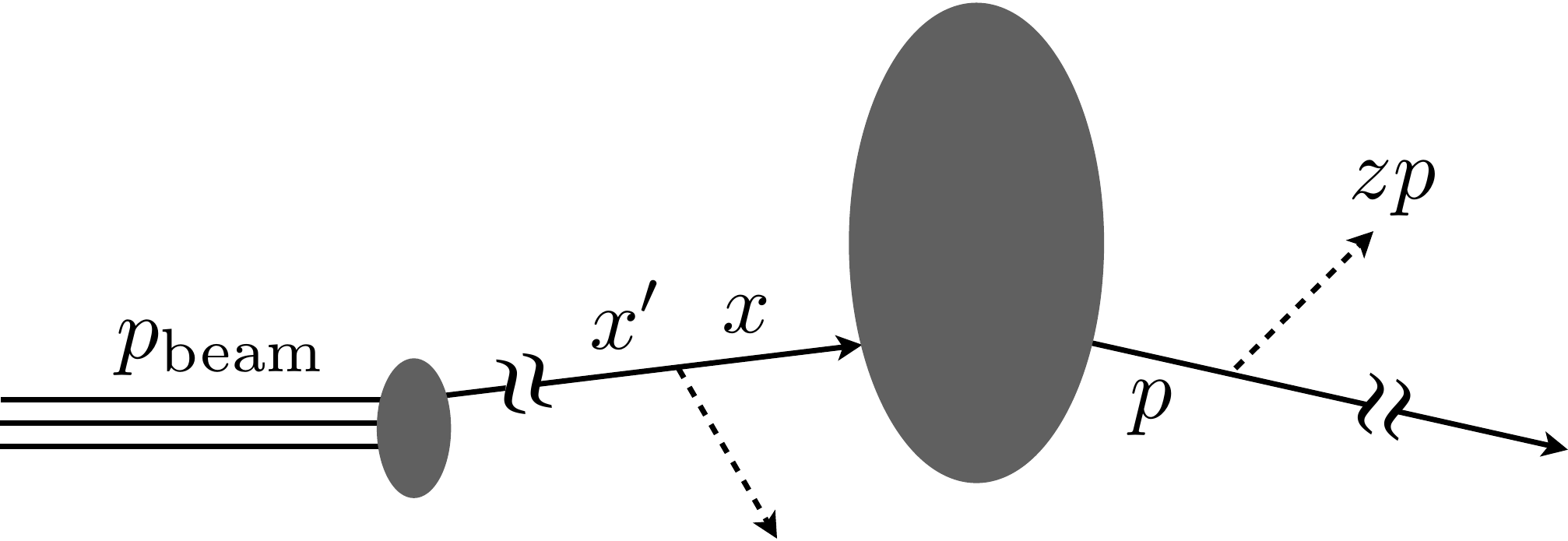}
\caption{{\footnotesize Schematic illustration of initial and final state radiations. The central blob shows a hard process, $p$ is the transverse momentum for a final state, $z$ is the energy fraction for the final state, and $x$ and $x'$ are the momentum fractions for an initial state and its parent parton.}}

\label{fig:ME}
\end{center}
\end{figure}

We derive evolution equations for generating functionals of final and initial states.  
We first start with the final state.  
We get following equations in the case that any resolved emission doesn't happen, namely for $n=1$, 
\begin{align}
\mathcal{R}_1^{(q, {\rm out})}(v,p,t) = v\DeT_q(p,t), \quad
\mathcal{R}_1^{(g, {\rm out})}(v,p,t) =   \DeT_g(p,t), \label{Rout1}
\end{align}
where $\DeT_i(p,t)$ is the Sudakov form factors which shows the probability that any resolved emission doesn't happen between the starting scale $t$ and a minimum resolved scale.  We define the form factors in Sec.~\ref{Sudakov_factors}.  
If a quark doesn't emit any resolved emission it forms one quark jet, so we need $v$ in front of $\DeT_q$. 
In the case that the resolved emission happens at least one time, namely for $n>1$, the modified jet rates have the following equation, 
\begin{align}
\mathcal{R}_n^{(i, {\rm out})}(v,p,t) 
&=  \sum_k \int_{p_0/p}^{1} \frac{dz}{z} \int_{t_0}^{t} \frac{dt'}{t'}
    \frac{\DeT_i(p,t)}{\DeT_i(p,t')} \GammaT_{i\to jk}(z,t') \label{Rout2}\\
    &\hspace{40pt} 
    \times \sum_{n_1+n_2=n}
    \mathcal{R}_{n_1}^{(j, {\rm out})}(v,p,t') \mathcal{R}_{n_2}^{(k, {\rm out})}(v,zp,t'), 
    \nonumber
\end{align}
where $\GammaT_i(z,t)=\alpha_s(z,t) P_i(z)/\pi$, 
$\alpha_s(z,t)$ and $P_i(z)$ are the running strong coupling and the splitting functions, 
$t_0$ is a given minimal scale,  
and $p_0$ is the minimum resolved transverse momentum which corresponds to the minimum $p_T$ cut for jets. 
We use the emission angle as the scale such as $t=\sqrt{1-\cos\theta}~(\simeq \theta/\sqrt{2})$, then $t_0=\sqrt{1-\cos R}$, where $R$ corresponds to the jet radius. 
As said above, we simply declare radiation whose transverse momentum and angle are greater than $p_0$ and $R$ as a jet in our calculation, and the jet algorithm dependence is beyond the scope of this paper. 
The ratio of Sudakov form factor shows the probability that the parton $i$ doesn't emit any resolved emissions between the scale $t$ and $t'$. 
From these equations, we get evolution equations for the generating functionals of final states \cite{Gerwick:2012fw,Bhattacherjee:2015psa} as, 
\begin{align}
\Phi_q(p,t) 
&= u v \DeT_q(p,t)
    + \int_{p_0/p}^{1} \frac{dz}{z} \int_{t_0}^{t} \frac{dt'}{t'}
    \frac{\DeT_q(p,t)}{\DeT_q(p,t')}
    \GammaT_q(z,t') \Phi_q(p,t') \Phi_g(zp,t'),  \label{genFq}\\
\Phi_g(p,t) 
&= u \DeT_G(p,t)
    + \int_{p_0/p}^{1} \frac{dz}{z} \int_{t_0}^{t} \frac{dt'}{t'}
	\frac{\DeT_G(p,t)}{\DeT_G(p,t')}
	\left[ \GammaT_g(z,t') \Phi_g(p,t') \Phi_g(zp,t') \right.   \label{genFg}\\
     & \hspace{68mm}
     \left. + n_f\GammaT_{gg}(z,t') \Phi_q(p,t') \Phi_q(zp,t')   \right]. \nonumber
\end{align}
where $\Phi_q$ and $\Phi_g$ are the generating functionals for quarks and gluons. 
For brevity we define the following logarithms,
\begin{alignat}{3}
\kappa &= \ln(p/p_0), ~~~~~~~~& \kappa'  &= \ln(zp/p_0),\\
\lambda &= \ln(t/t_0),                   & \lambda' &= \ln(t'/t_0). 
\end{alignat}  
With these variables, the equations in (11) and (12) are rewritten as,
\begin{align}
\Phi_q(\ka,\la) &= uv \DeT_q(\ka,\la)
	\exp \left[
	\int_0^{\la} d\la' \int_0^{\ka} d\ka'  \GammaT_q(z,t') \Phi_g (\ka' ,\la' )
	\right],  \label{Phi_q}\\
\Phi_g(\ka,\la) &= u \DeT_G(\ka,\la)
	\exp \left\{
	\int_0^{\la} d\la' \int_0^{\ka} d\ka'  \left[
		\GammaT_g(z,t') \Phi_g (\ka' ,\la' ) \right.\right. \label{Phi_g} \\
	&\hspace{52mm}\left.\left. 
		+n_f \GammaT_{\qq}(z,t') \frac{\Phi_q(\ka,\la')}{\Phi_g(\ka,\la')} \Phi_q (\ka' ,\la')
		\right]
	\right\}, \nonumber
\end{align}

We next derive generating functionals for initial states.  
For an initial state $i$, from the DGLAP equation, the normalized change of a parton density for $i$, in other words, the probability that $i$ emits an initial state radiation between $t'$ and $t'+dt'$ is,
\begin{align}
d\mathcal{P}(x,t') = \frac{d f_i(x,t')}{f_i(x,t')} 
&= \sum_k
	\frac{dt'}{t'} \int_{x}^1 \frac{dx'}{x'} \frac{\alpha_s}{\pi}
	P_{k \to ij} (z) \frac{f_k(x',t')}{f_i(x,t')}, 
\\
&= \sum_k
	\frac{dt'}{t'} \mathcal{P}_{k \to ij}(z,t'),
\end{align}
where $x$ and $x'$ are the momentum fractions for $i$ and its parent parton as shown in Fig.~\ref{fig:ME}, $f_i$ is the parton distribution function (PDF) for $i$, and $z=(x'-x)/x'$. 
In the case of initial state radiation, $\mathcal{R}_0^{(i, {\rm in})}$ is given by the probability that $i$ doesn't emit any resolved initial state radiations, 
\begin{align}
\mathcal{R}_0^{(i, {\rm in})}(v,x,t) = \exp\left(-\int_{t_0}^t d\mathcal{P}(x,t') \right) = \PiT_i(x,t),
\end{align}
where $\PiT_i$ is the Sudakov form factor for initial states.  In the case that a resolved emission happens at least one time, namely for $n>0$, the modified jet rates have the following relation, 
\begin{align}
\mathcal{R}_n^{(i, {\rm in})}(x,t)
&=  \sum_k \int_{t_0}^{t} \frac{dt'}{t'}
    \frac{\PiT_i(x,t)}{\PiT_i(x,t')} \mathcal{P}_{k\to ij}(z,t')\\
    &\hspace{40pt} 
    \times \sum_{n_1+n_2=n}
    \mathcal{R}_{n_1}^{(j, {\rm in})}(x',t') \mathcal{R}_{n_2}^{(k, {\rm out})}((x'-x)p_{\rm beam},t'). 
\end{align}
From the above equations and the definitions of the generating functionals, the evolution equation for the functional of initial states is given as, 
\begin{align}
\Psi_i(x,t)
&= \Pi_i(x,t)+\sum_k \int_{t_0}^t \frac{dt'}{t'} \int_x^1\frac{dx'}{zx'} \frac{\Pi _i(x,t)}{\Pi _i(x,t')} \label{genI} \\
	&\hspace{10mm}\times
	\frac{f_k(x',t)}{f_i(x,t)} \Gamma_{k\to ij}(z,t')
	\Psi _{k} (x',t') \Phi_j ((x'-x)E_{\rm beam},t'), 
	\nonumber
\end{align}
The logarithms $\ka$ and $\ka'$ are modified for initial states as,
\begin{align}
\kaB =\ln ( (1-x)p_{\rm beam}/p_0 ), \quad
\kaB'=\ln ( (x'-x)p_{\rm beam}/p_0 ).
\end{align}
In terms of these variables, the equations in (\ref{genI}) for quarks and gluons are,
\begin{align}
\Psi _q(\kaB,\la )&=\PiT _q (\kaB,\la ) 
	\exp \left[
	\int _0^{\la} d\la'   \int_0^{\kaB}  d\kaB'    
	\left\{
		\GammaT_{q}(z,t') \frac{f_q(x')}{f_q(x)}  \Phi_g (\kaB', \la' )
		\right.\right. \label{eq:PsiQ1} \\
		&\hspace{55mm} \left.\left. 
		+\GammaT_{\qq}(z,t') \frac{f_g(x')}{f_q(x)}  \frac{\Psi_g(x',t')}{\Psi_q(x,t')}   \Phi_q(\kaB', \la')
		\right\}
	\right], \nonumber \\
\Psi _g(\kaB,\la )&=\PiT _g (\kaB,\la ) 
	\exp \left[
	\int _0^{\la} d\la'   \int_0^{\kaB}  d\kaB'    
	\left\{
		\GammaT_{g}(z,t') \frac{f_g(x')}{f_g(x)}  \Phi_g (\kaB', \la' )
		\right.\right. \label{eq:PsiG1}\\
		&\hspace{48mm} \left.\left. 
		+\sum_q \GammaT_{gq}(z,t') \frac{f_q(x')}{f_g(x)}  \frac{\Psi_q(x',t')}{\Psi_g(x,t')}   \Phi_q(\kaB', \la')
		\right\}
	\right], \nonumber 
\end{align}
where we neglect the scale dependence on the ratio of PDF since the effect is negligible.\footnote{We fix the factorization scale for the PDF ratios to the hard scale, namely $t'=t$, in numerical calculations below.}

The splitting kernels are summarised as follows:
\begin{align}
\int \frac{dz}{z}\, \tilde{\Gamma}_i(z,t')
 & =\int dz\, \frac{\alpha_s(k_t^2)}{\pi} P_i(z) \\ 
 &\simeq \int \frac{dz}{z}\, (1-D) \times
 \begin{cases}
  a_q, & i=q, q\to qg, \\
  a_g, & i=g, g\to gg, \\
  \frac{a_{qq}}{n_f}z[z^2+(1-z)^2], & i=\qq, g\to q\bar{q}, \\
  a_q \frac{z}{2}\frac{1+z^2}{1-z}, & i=gq, q\to gq, 
\end{cases} \label{splitting_functions} \\
 &= \int \frac{dz}{z}\, (1-D)\times \Gamma_i(z).
\end{align}
We use the relative transverse momentum $k_t$ as the scale of the strong coupling, and expand the coupling at a minimal $k_t$ with the 1-loop beta function, 
\begin{align}
&\alpha_s(k_t^2) = \overline{\alpha}_s(1-D), \quad 
\overline{\alpha}_s = \alpha_s(2p_0^2 t_0^2), \quad 
a=2 \overline{\alpha}_s b_0, \\
& D = 
\begin{cases}
  (a\ka'+a\la')/(1+a\ka'+a\la'), & \text{for final states}, \\
  (a\kaB'+a\la')/(1+a\kaB'+a\la'), & \text{for initial states}, \\
\end{cases}
\end{align}
where we employed the following expression for the transverse momentum, 
\begin{align}
& k_t^2 = \begin{cases}
  2 z^2 p^2                         t'^2 & \text{for final states}, \\
  2 (x'-x)^2p_{\rm beam}^2 t'^2 & \text{for initial states}. 
\end{cases}
\end{align}
The coefficients are $a_{q,g}=2C_{F,A}\overline{\alpha}_s/\pi$ and $a_{\qq}=n_f T_R\overline{\alpha}_s/\pi$ for final states, and we remove the factor 2 in $a_g$ for initial states because the soft singularity for $z\to 1$ in the gluon splitting function $P_g(z)$ is suppressed by the gluon PDF  $f_g(x')$.  
The number of active flavors is given by $n_f$ and it is set to 5 in numerical calculations below.
The non-tilde splitting kernel $\Gamma_i(z)$ is given by removing the running effect of $\alpha_s$ from $\GammaT_i(z,t')$.

\subsubsection{Sudakov form factors}
\label{Sudakov_factors}
The Sudakov form factors for final states are defined as,
\begin{align}
\DeT_i(\ka,\la)       &=  \exp \left( -\int_0^{\ka}d\ka' \int_0^{\la}d\la'       \GammaT_i(z,t')        \right), \quad i\in\{q,g\}\\ 
\DeT_{\qq}(\ka,\la) &=  \exp \left( -\int_0^{\ka}d\ka' \int_0^{\la}d\la' n_f \GammaT_{\qq}(z,t') \right),\\
\DeT_{G}(\ka,\la)   &=  \DeT_{g}(\ka,\la) \DeT_{\qq}(\ka,\la).
\end{align}
Sudakov form factors which are evaluated by neglecting the running effect of $\alpha_s$ are given as,
\begin{align}
\De_{i}(\ka,\la)     &=  \exp ( -a_{i} \ka\la ), \quad i\in\{q,g\}, \label{eq:Delta_i}\\
\De_{\qq}(\ka,\la) &= \exp ( -a_{\qq} c_{\qq}\la ), \\
\De_G(\ka,\la) &= \De_{g}(\ka,\la) \De_{\qq}(\ka,\la),
\end{align}
where $c_{\qq}=\frac{2}{3}(1-e^{-3\ka})+e^{-2\ka}-e^{-\ka} \sim \frac{2}{3}$.
We can see the structure of leading (or double) logarithms (LL) in $\De_{q}$ and $\De_{g}$, and single logarithms in $\De_{\qq}$.

\begin{figure}[!t]
\begin{center}
\includegraphics[width=7.5cm, bb=0 0 350 201]{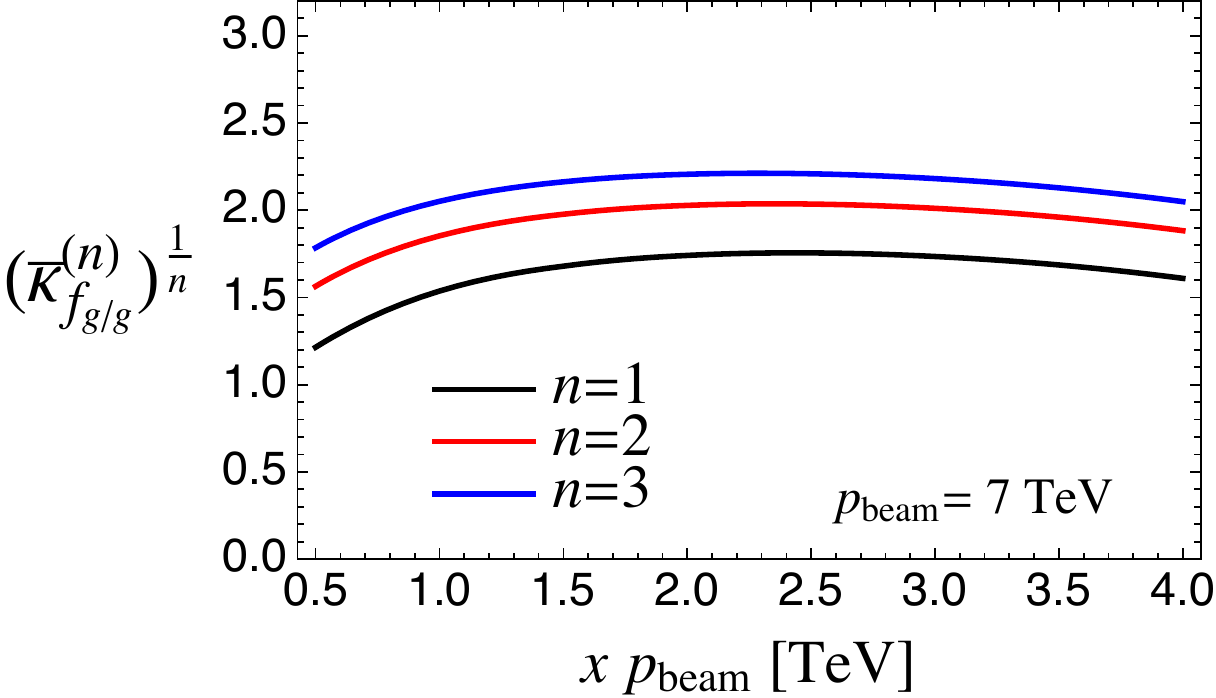} \quad
\includegraphics[width=7.5cm, bb=0 0 350 201]{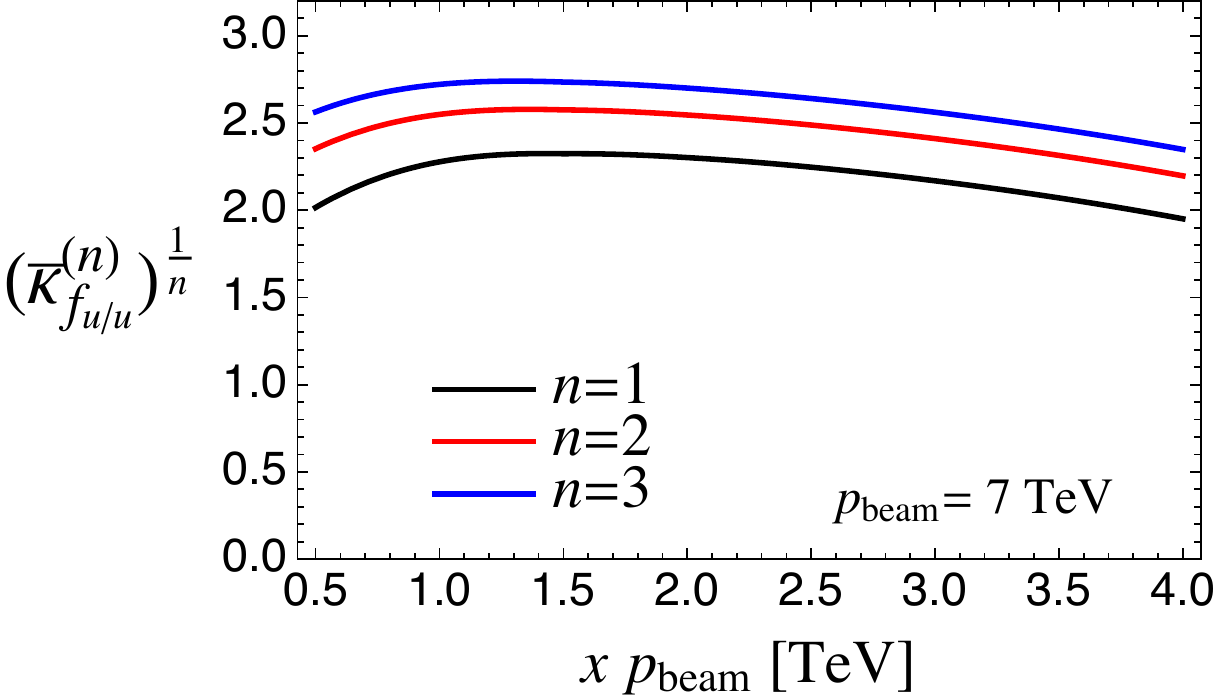} \\
\includegraphics[width=7.0cm, bb=0 0 350 215]{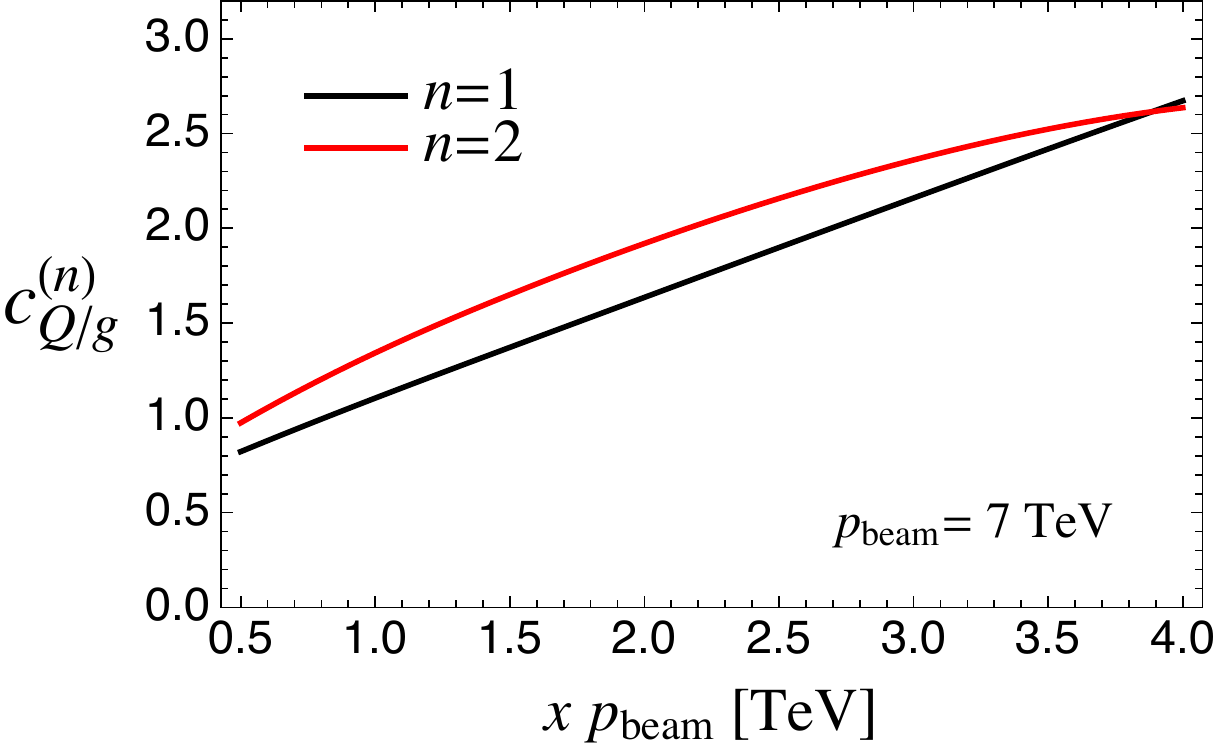} \quad
\includegraphics[width=7.0cm, bb=0 0 350 218]{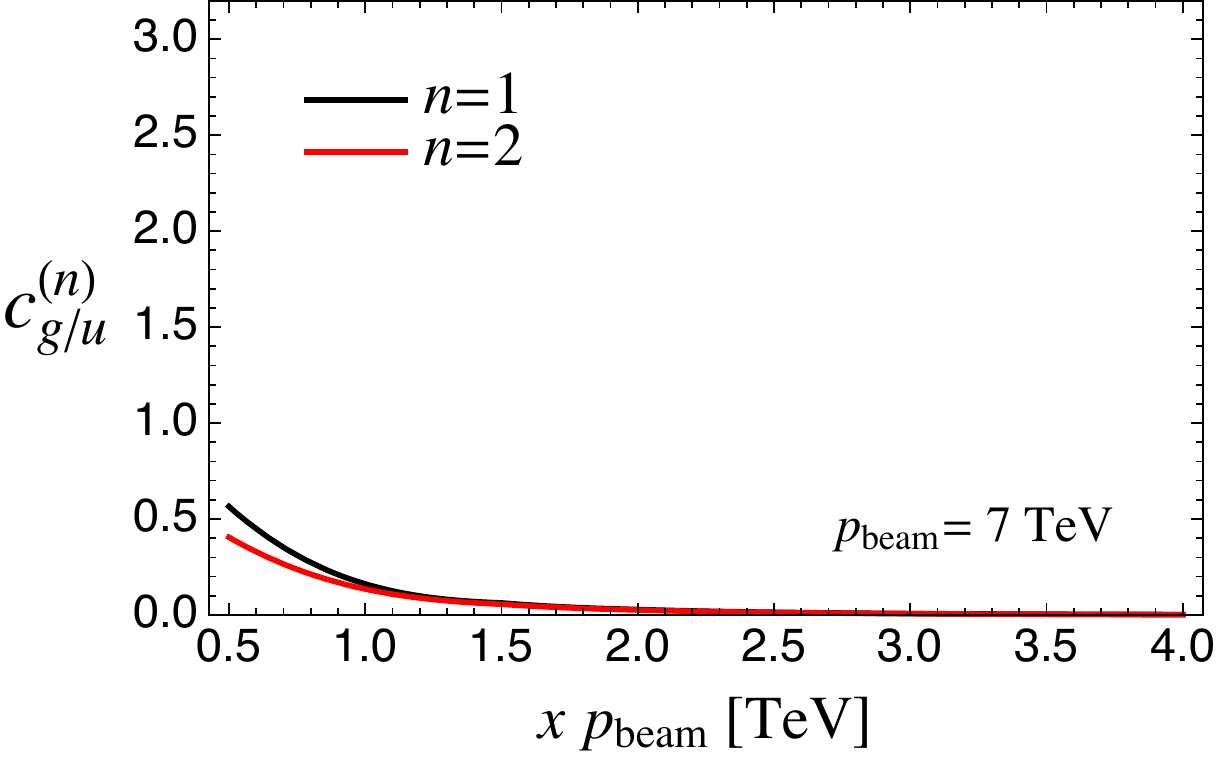} 
\caption{{\footnotesize 
The $x$-dependence of coefficients in Eqs.~(\ref{eq:kapbar})-(\ref{cQg}). 
The vertical axis shows $xp_{\rm beam}$ which is the energy of an initial state.
CTEQ6L1 is used in the calculations.
}}
\label{kappa_c_coe}
\end{center}
\end{figure}

For initial states, the Sudakov factors are defined as,
\begin{align}
\PiT_{i}(\kaB,\la) &= \PiT_{i,1}(\kaB,\la) \PiT_{i,2}(\kaB,\la), \quad i\in\{q,g\}, \\
\PiT_{i,1}(\kaB,\la)  &= \exp \left(   -\int_0^{\kaB}d\kaB' \int_0^{\la}d\la'         \frac{f_i(x')}{f_i(x)}   \GammaT_{i}(z,t')   \right) ,\\
\PiT_{q,2}(\kaB,\la) &= \exp \left(   -\int_0^{\kaB}d\kaB' \int_0^{\la}d\la'         \frac{f_g(x')}{f_q(x)}   \GammaT_{\qq}(z,t')    \right), \\
\PiT_{g,2}(\kaB,\la) &= \exp \left(   -\int_0^{\kaB}d\kaB' \int_0^{\la}d\la'  \sum_q  \frac{f_q(x')}{f_g(x)}   \GammaT_{gq}(z,t')    \right), 
\end{align}
Neglecting the running of $\alpha_s$, we get
\begin{align}
\Pi_{i}(\kaB,\la) &= \Pi_{i,1}(\kaB,\la) \Pi_{i,2}(\kaB,\la), \quad i\in\{q,g\}, \\
\Pi_{i,1}(\kaB,\la)  &= \exp(-a_i \kaB^{(1)}_{f_{i/i}}\la), \\
\Pi_{q,2}(\kaB,\la) &= \exp(- c^{(1)}_{g/q}  \frac{a_{\qq}}{n_f}  \la), \\
\Pi_{g,2}(\kaB,\la) &= \exp(- c^{(1)}_{Q/g}  a_q \la).
\end{align}
We define a functionalized $\kaB$ with a function $f$ as,
\begin{align}
\kaB^{(n)}_f = n \int_0^{\kaB} d\kaB' \,  \kaB'^{n-1}   f(\kaB') , \label{eq:kapbar}
\end{align}
where $f_{i/i} (\kaB) = f_i(x')/f_i(x)$. 
For an identity function $I$, we can find a simple relation, $\kaB^{(n)}_I = \kaB^{n}$.  
The two coefficients are given as,
\begin{align}
c^{(n)}_{g/q}
	&=   \frac{n_f}{a_{\qq}}  \int_0^{\kaB}  d\kaB' \frac{            f_g(x')}{f_q(x)}   \Gamma_{\qq}(z) \times \kaB'^{n-1} , \label{cgq}\\
c^{(n)}_{Q/g}
	&=   \frac{1}{a_q}          \int_0^{\kaB}  d\kaB' \frac{\sum_q f_q(x')}{f_g(x)}    \Gamma_{gq}(z) \times \kaB'^{n-1} , \label{cQg}
\end{align}
where $\sum_q$ runs over all active quarks and anti-quarks.
Fig.~\ref{kappa_c_coe} shows $x$-dependence of the coefficients in Eqs.~(\ref{eq:kapbar})-(\ref{cQg}). 
The vertical axis shows $xp_{\rm beam}$ which is the energy of an initial state.
Valence quarks become dominant at large $x$, therefore, $c_{Q/g}^{(n)}$ becomes bigger and $c_{g/u}^{(n)}$ becomes smaller as $xp_{\rm beam}$ increases. 
We adopt the CTEQ6L1 PDF \cite{Pumplin:2002vw} in the calculations with the help of a PDF parser package, {\tt ManeParse 2.0} \cite{Clark:2016jgm}.

\subsection{Primary structure of functionals}
\label{sec:app1}
Since the largest contribution to the $t$ integration in Eqs.~(\ref{genFq}) and (\ref{genFg}) comes from the region $t' \sim t$,  we use approximations $\Phi_q(\la') \sim \Phi_q(t_0)=uv$ and $\Phi_g(\la') \sim \Phi_g(t_0)=u$ to see primary structures of the functionals~\cite{Gerwick:2012hq}.\footnote{In the approximations,  the functional ratio in Eq.~(\ref{Phi_g}) takes a form $\Phi_q(\ka,\la')/\Phi_g(\ka,\la') \sim uv/u$, which causes unphysical terms, $u^n v^m(n<m)$.  We remove such terms by hand, which causes a unitarity violation, namely $\Phi |_{u=1,v=1}\neq 1$.  However the violation is very tiny, so we keep using the approximations.  Note that the unitarity violation rate is $1-\Phi |_{u=1,v=1} \simeq (0.1$-$0.4)\%$ for numerical results in this paper.}
In these approximations, subsequent emissions from a low-scale parton are prohibited. 
We correct the absence of the subsequent emissions effect in the next section. 
Applying the approximations in the evolution equation (\ref{Phi_q}) and (\ref{Phi_g}) and 
neglecting the running of $\alpha_s$, we get 
\begin{align}
\Phi_q^{(\LL+\qq)} &= uv \Delta_q
	\exp \left(
	\int_0^{\la} d\la' \int_0^{\ka } d\ka'  \Gamma_q(z) u
	\right),\\
	&= uv \Delta_q^{1-u}, \label{eq:Delta_q_LL}\\
\Phi_g^{(\LL+\qq)} &= u \Delta_G
	\exp \left\{
	\int_0^{\la} d\la' \int_0^{\ka } d\ka'  \left[
		\Gamma_g(z) u +n_f \Gamma_{\qq}(z) uv^2
		\right]
	\right\}, \\
	&= u \Delta_g^{1-u} \Delta_{\qq}^{1-uv^2}, \label{eq:Delta_g_LL}
\end{align}
The terms $\Delta_{q,g}^{1-u} (\propto \sum_n u^n \cdot \asB^n L^{2n}/n!)$ which come from the integration of leading splitting kernels $\Gamma_{q,g}$ are involved in the increment of the number of gluon jets with the double logarithmic coefficients $\asB^n L^{2n}$, where $L$ is proportional to $\ka$ or $\la$.  
The term $\Delta_{\qq}^{1-uv^2} (\propto \sum_n u^nv^{2n} \cdot \asB^n L^n/n!)$ contains $v$, so it is involved in the increment of the number of quark jets with the single logarithmic coefficients $\asB^n L^n$.  
Since the enhancement term of gluon jets  has more logarithmic enhancement, the increment of gluon jets is larger than that of quark jets. 
For $\Phi_q^{(\LL+\qq)}$, the functional doesn't contain the $\Delta_{\qq}$ term, so the number of quark jets doesn't increase in this approximation.

Regarding the evolution equation for initial states in Eqs.~(\ref{eq:PsiQ1}) and (\ref{eq:PsiG1}),
we adopt the approximation $\Psi_i(\la') \sim \Psi_i(t_0) = 1$ as with the case of final states.  
The primary structures of the functionals is represented by,
\begin{align}
\Psi_i^{(\LL+\qq)} &= \Pi_{i,1}^{1-u}   \Pi_{i,2}^{1-uv},  \quad  i\in \{q,g\}.
\end{align}
The structure of Sudakov form factors are,
\begin{align}
\Pi_{i,1}^{1-u} &\propto \sum_n u^n        \cdot \left( a_i \kaB_{f_{i/i}}^{(1)}\la \right)^n /n!, \\
\Pi_{q,2}^{1-u} &\propto \sum_n u^nv^n \cdot \left( \frac{a_{\qq}}{n_f} c^{(1)}_{g/q} \la \right)^n /n!, \\
\Pi_{g,2}^{1-u} &\propto \sum_n u^nv^n \cdot \left( a_q c^{(1)}_{Q/g} \la \right)^n /n!. 
\end{align}
Since the leading Sudakov $\Pi_{i,1}^{1-u}$ doesn't contain $v$, it's not involved in the increment of the number of quark jets. 
On the other hands, the sub-leading Sudakov $\Pi_{i,2}^{1-u}$ increase the number of quark jets since it contains $v$. 
In Fig.~\ref{kappa_c_coe}, we noticed that the coefficient $\kaB_{f_{i/i}}^{(1)}$ is basically larger than $c^{(1)}_{i/j}$ since the former is given by the integral of the splitting kernels having the soft-singularity terms. 
Therefore the increment of gluon jets is basically larger than that of quark jets as with the case of final state radiations. 
However, quark jets are easier to be radiated from gluon initial state at high energy since $c^{(1)}_{Q/g}$ gets bigger as the hard scale increases.

\subsection{Corrections by subsequent emissions}
\label{sec:app2}
We add a correction to the generating functionals evaluated in the last section. 
In the previous section, we use the approximation $\Phi_g(\ka',\la')=u$. 
This means that emissions by soft gluons which carry the energy fraction $z$ are neglected. 
Consequently, a quark in final states doesn't make quark jets more than one due to the absence of $g\to \qq$. 
We modify the approximation in order to take into account the radiations by the soft gluons as $\Phi_g(\ka',\la') \simeq \Phi_g^{(\LL+\qq)}(\ka',\la')$. 
We also modify the approximation for the ratio of functionals.  
We found that the primary structure of the functionals has the exponential suppression $\exp(-c \ka)$, thus the precision of approximation employed in the previous section, i.e., $\Phi_q(\ka,\la')/\Phi_g(\ka,\la')=1$ gets worse as increasing the energy scale $\ka$. 
In order to take into account the energy scaling of the ratio, we employ a new approximation $\Phi_q(\ka,\la')/\Phi_g(\ka,\la') \simeq \Phi_q^{(\LL+\qq)}(\ka,\la')/\Phi_g^{(\LL+\qq)}(\ka,\la')$. 
Applying these approximations in the evolution equation (\ref{Phi_q}) and (\ref{Phi_g}) and 
neglecting the running of $\alpha_s$, we get 
\begin{align}
\Phi_q^{(\LL+\qq+\sub)} &= uv \Delta_q
	\exp \left[
	\int_0^{\la} d\la' \int_0^{\ka } d\ka'  \Gamma_q(z,t')
		u \Delta_g^{1-u}(\ka',\la') \Delta_{\qq}^{1-uv^2}(\ka',\la')
	\right],\\
	&= \Phi_q^{(\LL+\qq)} \times  \exp(S_q), \label{eq:S_q}\\
\Phi_g^{(\LL+\qq+\sub)} &= u \Delta_G
	\exp \left\{
	\int_0^{\la} d\la' \int_0^{\ka } d\ka'  \left[
		\Gamma_g(z,t') 
			u \Delta_g^{1-u}(\ka',\la') \Delta_{\qq}^{1-uv^2}(\ka',\la') \right.\right. \\ 
		& \hspace{44mm} \left.\left.
		+n_f \Gamma_{\qq}(z,t') 
			\frac{ \Delta_q^{1-u}(\ka,\la') }{ \Delta_g^{1-u}(\ka,\la') \Delta_{\qq}^{1-uv^2}(\ka,\la') }
			uv^2
		\right]
	\right\}, \nonumber \\
	&= \Phi_g^{(\LL+\qq)} \times  \exp(S_g)  \exp(S').
\end{align}
The exponential factor $\exp(S_i)$ stems from the modification for the soft-gluon generating functional.  
In other words, the term arises from activating subsequent emissions by soft-gluons. 
The term $\exp(S')$ stems from the modification for the functional ratio. 
Their full formulae are shown in Appendix~\ref{app:formulae}. 
Leading terms for the exponents are as follows:
\begin{align}
S_i &\sim  -u \ln\Delta_i \left[  -\frac{(1-u)a_g \ka\la}{4}  -\frac{(1-uv^2)a_{\qq} \la}{2}  \right], \label{eq:S_i}\\
S'   &\sim  -uv^2 \ln\Delta_{\qq}  \frac{w}{2}, \label{eq:S'}\\
w &= (1-u)(a_g-a_q)\ka\la    +   (1-uv^2)c_{\qq}a_{\qq} \la.
\end{align}
In Eq.~(\ref{eq:S_i}), the double logarithmic term $-u\ln\Delta_i$ comes from the soft-gluon emission $i \to i+g$.  The double and single logarithms in the square brackets comes from the subsequent emissions by the soft-gluon, and their fractional factors arise from the integrals of the ordering variables $\ka'$ and/or $\la'$. 
In Eq.~(\ref{eq:S'}), $w/2$ comes from the correction to the functional ratio.

As with the case of final states, we adopt the approximations $\Phi_i(\la') \sim \Phi_i^{(\LL+\qq)}(\la')$ and $\Psi_i(t') \sim \Psi_i^{(\LL+\qq)}(t')$ for the evolution equations for initial states in Eqs.~(\ref{eq:PsiQ1}) and (\ref{eq:PsiG1}), then we get 
\begin{align}
\Psi_i^{(\LL+\qq+\sub)} &= \Psi_i^{ (\LL+\qq)} \times  \exp( S_i[f_{i/i}] )  \exp(S_i'). 
\end{align}
For an analytic function $G = \sum_n c_n \ka^n$, we define a functionalized one, $G[f]$, as,
\begin{align}
G[f] = \sum_n c_n \kaB^{(n)}_f. \label{functionalized_G}
\end{align}
The full formula of $S'_i$ is shown in Appendix \ref{app:formulae}, and the leading terms are,
\begin{gather}
S'_q  \sim  -uv \ln\Pi_{q,2} \frac{-\overline{w}}{2}, \\
S'_g  \sim  -uv \ln\Pi_{g,2} \frac{ \overline{w}}{2},  \\
\overline{w} =
  (1-u  )(a_g\kaB_{g/g}^{(1)}  - a_q                     \kaB_{q/q}^{(1)}) \la 
+(1-uv)(a_q c_{Q/g}^{(1)}     -  \frac{a_{\qq}}{n_f}   c_{g/q}^{(1)} ) \la, \label{eq:wbar}
\end{gather}
Regarding $S'_q$ and $S'_g$, $-uv \ln\Pi_{i,2}$ comes from the sub-leading splitting kernels for $g\to \qq$ and $g\to gq$ in Eqs~(\ref{eq:PsiQ1}) and (\ref{eq:PsiG1}). 
The following factors $\pm \overline{w}/2$ come from the corrections to the functional ratios. 
The sign of $\overline{w}/2$ are opposite since the numerators and denominators for the functional ratios are opposite.

\subsection{$\alpha_s$ running correction}
\label{sec:app3}
Finally, we consider the running effect of $\alpha_s$. 
Taking into account the effect of subsequent emission discussed in the previous section and using the Sudakov factors and the splitting functions with a tilde, i.e., $\DeT_i$ and $\GammaT_i$, we get the following equations for final states, 
\begin{align}
\Phi_q^{(\LL+\qq+\sub+\delta\alpha_s)} &= uv \DeT_q
	\exp \left[
	\int_0^{\la} d\la' \int_0^{\ka } d\ka'  \GammaT_q(z,t')
		u \De_g^{1-u}(\ka',\la') \De_{\qq}^{1-uv^2}(\ka',\la')
	\right],\\
	&= \Phi_q^{(\LL+\qq+\sub)} \times  \exp(\tilde{S}_q)  \exp(T_q), \\
\Phi_g^{(\LL+\qq+\sub+\delta\alpha_s)} &= u \DeT_G
	\exp \left\{
	\int_0^{\la} d\la' \int_0^{\ka } d\ka'  \left[
		\GammaT_g(z,t') 
			u \De_g^{1-u}(\ka',\la') \De_{\qq}^{1-uv^2}(\ka',\la') \right.\right. \\ 
		& \hspace{35mm} \left.\left.
		+n_f \GammaT_{\qq}(z,t') 
			\frac{ \De_q^{1-u}(\ka,\la') }{ \De_g^{1-u}(\ka,\la') \De_{\qq}^{1-uv^2}(\ka,\la') }
			uv^2
		\right]
	\right\}, \nonumber \\
	&= \Phi_g^{(\LL+\qq+\sub)} \times  \exp(\tilde{S}_g)  \exp(T_g)  \exp(\tilde{S}')  \exp(T'). 
\end{align}
The full formulae for the exponential factors are shown in Appendix~\ref{app:formulae} and their leading terms are,
\begin{align}
\tilde{S}_i &\sim u \ln \Delta_i   \left[   -\frac{1}{6}(1-u)a_g\ka\la   \cdot a       (\ka              + \la                       )
						         -(1-uv^2)c_{\qq}a_{\qq}\la \cdot a \left(\frac{\ka}{4} + \frac{\la}{3} \right)   \right], \\
\tilde{S}'  &\sim  uv^2 \ln\Delta_{\qq} \cdot w \cdot  a \left(\frac{2\ka}{3} + \frac{\la}{3} \right), \\
T_i  & = (1-u)\ln(\DeT_i/\De_i) 
	  \sim  -(1-u) \ln\Delta_i  \cdot   a \frac{\ka + \la}{2},  \\
T'    & = (1-uv^2)\ln(\DeT_{\qq}/\De_{\qq}) 
	  \sim  -(1-uv^2)  \ln\Delta_{\qq}  \cdot  a \left(  \ka + \frac{\la}{2}  \right).
\end{align}
The two exponents $\tilde{S}_i$ and $\tilde{S}'$ are the $\alpha_s$ correction for $S_i$ and $S'$, and $e^{T_i}$ and $e^{T'}$ are the corrections for $\Delta_i$ and $\Delta_{\qq}$.

The $\alpha_s$ correction for the generating functionals of initial states are
\begin{align}
\Psi_i^{(\LL+\qq+\sub+\delta\alpha_s)} 
	&= \Psi_i^{(\LL+\qq+\sub)}  \times  \exp(\tilde{S}_i[f_{i/i}])  \exp(T_i[f_{i/i}])   \exp(\tilde{S}_i')  \exp(T_i'), 
\end{align}
where $\tilde{S}_i[f_{i/i}]$ and $T_i[f_{i/i}]$ are the functionalized $\tilde{S}_i$ and $T_i$, and 
please check the definition of functionalized in Eq.~(\ref{functionalized_G}).
The full formulae for the exponents $\tilde{S}_i'$ and $T_i'$ are shown in Appendix~\ref{app:formulae}, and their leading terms are as follows: 
\begin{align}
\tilde{S}'_q  &\sim   uv \ln\Pi_{q,2}   \cdot  w  \cdot a\left(  c_{g/q }^{(1)}  +  \frac{\la}{3}  \right), \\
\tilde{S}'_g  &\sim   uv \ln\Pi_{g,2}   \cdot  w  \cdot a\left(  c_{Q/g}^{(1)}  +  \frac{\la}{3}  \right), \\
T'_q  &\sim  -(1-uv) \ln\Pi_{q,2}  \cdot   a\,  c_{g/q}^{(2)}/c_{g/q}^{(1)}, \\
T'_g  &\sim  -(1-uv) \ln\Pi_{g,2}  \cdot   a\,  c_{Q/g}^{(2)}/c_{Q/g}^{(1)}.
\end{align}

\section{Numerical results}
\label{sec:Numerical-results}
\subsection{Number of quark jets}
\label{Number_of_quark_jets}
We evaluate the quark jet rate for a given Born configuration, $i_1i_2\to f_1f_2$. 
A generating functional for the configuration is given by,
\begin{align}
\bold{\Phi}_{i_1 i_2 \to f_1 f_2} = 
\Psi_{i_1}(x_1,t_{i_1})\Psi_{i_2}(x_2,t_{i_2})
\Phi_{f_1}(p_{f_1}, t_{f_1})\Phi_{f_2}(p_{f_2}, t_{f_2}).
\end{align}
We assume that the two final states scatter in the central region, which tends to occur at high energy, and set as
$\hat{p}_T = p_{f_1}=p_{f_2}=x_1 p_{\rm beam}=x_2 p_{\rm beam}$, where the proton beam energy is set to $p_{\rm beam}=7$ TeV.  
The starting scale is set to the maximal one allowed kinematically, namely $t_{\rm max}=\sqrt{2}$.

We calculate the number of quark jets for events in which $\Njets$ jets are contained. 
The expected value for the number is given by,
\begin{align}
\langle N_{\text{quark-jets}} \rangle =
	\frac
	{\sum_{m=0}^n m \, R_{n,m}^{(i_1 i_2 \to f_1 f_2)}}
	{R_{n}^{(i_1 i_2 \to f_1 f_2)}}, \quad n = \Njets,  \label{eq:nqjet_ave}
\end{align}
where the  jet rates and quark jet rates for $i_1 i_2 \to f_1 f_2$ are given as,
\begin{align}
R_{n}^{(i_1 i_2 \to f_1 f_2)}  &= 
	\left. \frac{1}{n!} \frac{\partial^n}{\partial u^n}
	\bold{\Phi}_{i_1 i_2 \to f_1 f_2} \right\vert_{u=0,v=1}, \\
R_{n,m}^{(i_1 i_2 \to f_1 f_2)}  &= 
	\left. \frac{1}{n! m!} \frac{\partial^n}{\partial u^n} \frac{\partial^m}{\partial v^m} 
	\bold{\Phi}_{i_1 i_2 \to f_1 f_2} \right\vert_{u=v=0}. 
\end{align}

\begin{figure}[!t]
\begin{center}
\includegraphics[width=0.32\linewidth, bb=0 0 300 230]{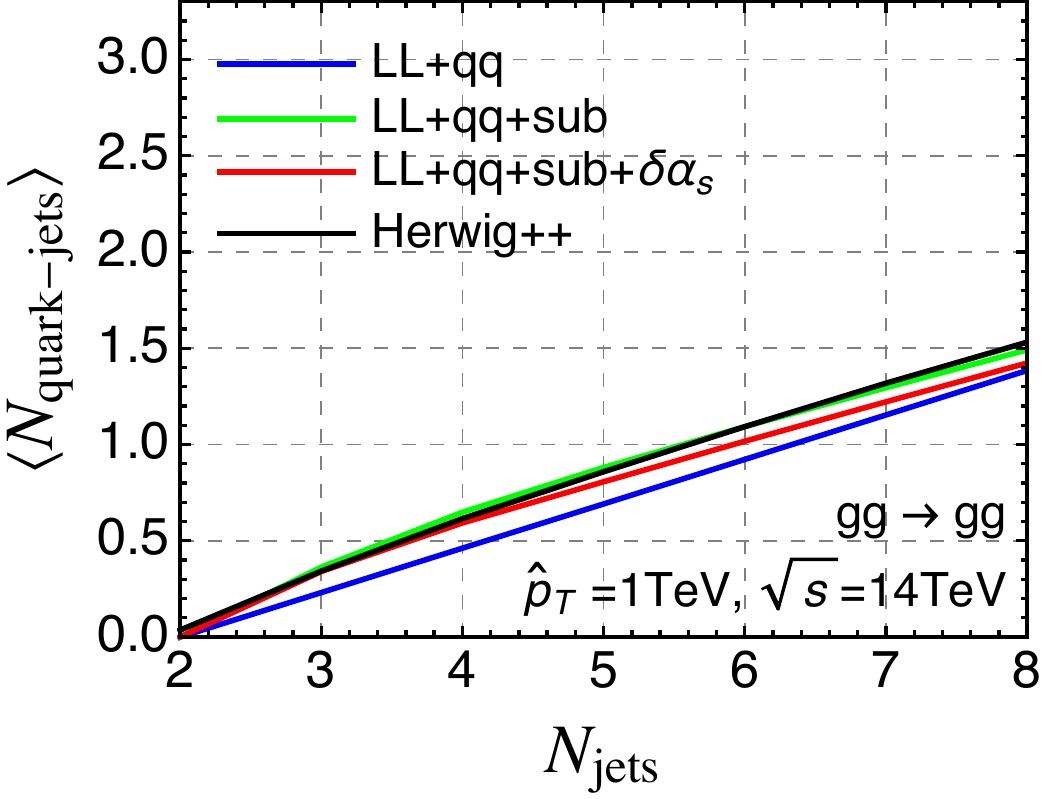}
\includegraphics[width=0.32\linewidth, bb=0 0 300 230]{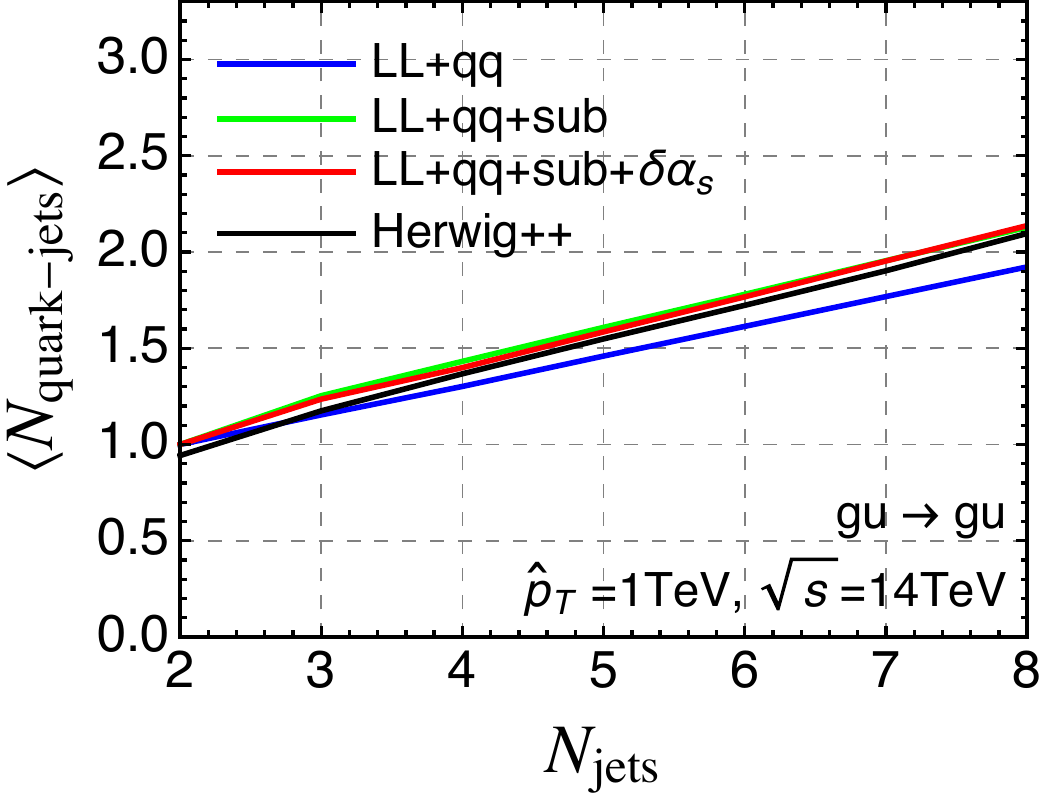}
\includegraphics[width=0.32\linewidth, bb=0 0 300 230]{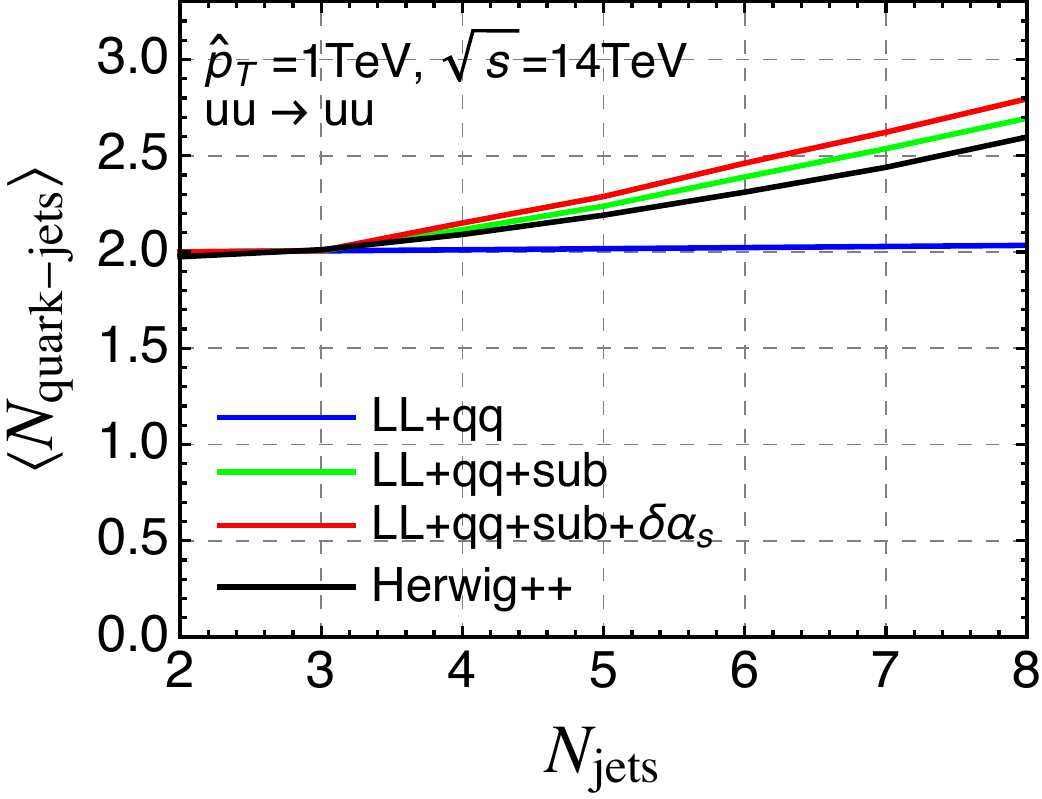}
\caption{{\footnotesize 
Expected values of the number of quark jet for events in which $\Njets$ jets are contained.  The results for $gg\to gg$ (left), $gu\to gu$ (center) and $uu\to uu$ (right) are shown.
In the calculation, $p_0=50$ GeV and $R=0.4$ are used. 
}}
\label{fig:Nq_all}
\end{center}
\end{figure}

In Fig.~\ref{fig:Nq_all}, the results for $gg\to gg$ (left), $gu\to gu$ (center) and $uu\to uu$ (right) are shown. 
In the calculation, $p_0=50$ GeV and $R=0.4$ are used. 
We set the parton transverse momentum as $\hat{p}_T = 1$ TeV. 
The blue, green and red curves are analytical results using the functionals labeled by $(\LL+\qq)$, $(\LL+\qq+\sub)$ and $(\LL+\qq+\sub+\delta\alpha_s)$. 
The black curves show Monte-Carlo predictions given by {\tt Herwig++} \cite{Bahr:2008pv}.\footnote{Our calculation employs a traditional angular-ordered parton shower formalism with $1\to 2$ splitting kernels. We compare our calculation with {\tt Herwig++} implementing the same formalism and the splitting kernels to check the consistency of our analytical results.} 
Hadronization is turned off and the generated partons are clustered by anti-$k_T$ algorithm \cite{Cacciari:2008gp}.\footnote{
As mentioned in Sec.~\ref{sec:Quark-jet-rates}, we will be able to measure the $\Njets$ dependence on $\langle N_\text{quark-jets} \rangle$ using the hadron-level definition of quark jets as in Ref.~\cite{Komiske:2018vkc}. 
}
We define a jet flavor from jet constituents. 
In the definition, it is necessary to consider the IR-unsafety of the jet flavor caused by soft gluon decaying into $q\bar{q}$ as referred to in Ref.~\cite{Banfi:2006hf}. 
We take into account the decay effect and details of estimating the number of quark jets in Monte-Carlo samples are summarized in Appendix~\ref{app:qjet_def}. 
We can see the analytic results including the terms of the subsequent emissions agree with the Monte-Carlo results.

The primary structure of functionals has the form with constants, 
\begin{align}
{\bf \Phi} \propto u^2 v^{m_0} e^{A u \la} e^{B uv \la} e^{C uv^2 \la}, \label{ref:mainStr}
\end{align}
where $A$, $B$ and $C$ are constants for a Born configuration.  
The expected value for the functional is estimated as,
\begin{align}
\langle N_\text{quark-jets} \rangle \simeq \frac{B+2C}{A+B+C}(\Njets-2) + m_0, 
	\label{eq:nqjet_inc}
\end{align}
where $m_0$ is the number of quarks in final states for a targeted Born configuration.
As increasing the coefficients $B$ and $C$ related to $v$, the number of quark jets increases. 
The three coefficients and the initial number of quark for three configurations $gg\to gg$, $gu\to gu$, and $uu\to uu$ are shown in Table~\ref{table:Nqjet}.
When we neglect subsequent emissions, the increase of quark jets for $uu\to uu$ is tiny because it is caused by only $B$ and the coefficient is much smaller than other coefficients as shown in Fig.~\ref{kappa_c_coe}. 
The main cause of the increase of quark jets for $uu\to uu$ stems from $\exp(S_q)$ which is related to subsequent emissions, and the lowest order at which $v$ appears is $\mathcal{O}(u^4v^4)$, therefore, the number of quark jets begins to increase clearly from $\Njets=4$.

You can also see auxiliary plots in Appendix \ref{app:ISR_FSR} where only initial or final state radiations are taken into account.

\begin{table}[!t]
  \centering
  \begin{tabular}{c|c|c|c|c}
    \hline
        $i_1i_2\to f_1f_2$  & $m_0$  
        		& $A$   
		& $B$   
		& $C$  \\
    \hline 
    $gg\to gg$  & 0  
    		& $2a_g(\ka+\kaB^{(1)}_{f_{g/g}})$  
		& $2c^{(1)}_{Q/g}a_q$  
		& $2c_{\qq}a_{\qq}$ \\
    \hline 
    $gu\to gu$  & 1  
    		& $a_g(\ka+\kaB^{(1)}_{f_{g/g}})+a_q(\ka+\kaB^{(1)}_{f_{q/q}})$  
		& $c^{(1)}_{Q/g}a_q + c^{(1)}_{g/u}\frac{a_{\qq}}{n_f}$  
		& $c_{\qq}a_{\qq}$ \\
    \hline 
    $uu\to uu$  & 2  
    		& $2a_q(\ka+\kaB^{(1)}_{f_{u/u}})$  
		& $2c^{(1)}_{g/u}\frac{a_{\qq}}{n_f}$  
		& 0 \\
    \hline
  \end{tabular}
  \caption{{\footnotesize
  Coefficients in Eq.~(\ref{ref:mainStr}) for three configurations $gg\to gg$, $gu\to gu$, and $uu\to uu$.
  }}
  \label{table:Nqjet}
\end{table}

\subsection{Expected improvement by the quark/gluon discrimination}
\label{sec:Expected_improvement}
In this subsection, we connect the knowledge obtained so far and BSM searches at the LHC. 
In Sec.~\ref{Number_of_quark_jets}, we can see that coefficient $A$ in Table \ref{table:Nqjet} which is related to the increment of gluon jets is basically bigger than $B$ and $C$. 
This means that many of QCD multi-jet background is composed of gluon jets and few quark jets stemming from the valence quarks. 
So, we expect to get a large improvement in the separation between QCD multi-jet backgrounds and signals containing many quark jets by using the quark/gluon tagging.

We estimate the improvement of signal-to-background ratio $(S/B)$ for such signals using the analytical results in Sec.~\ref{sec:Quark-jet-rates} and Monte-Carlo results. 
We introduce an improvement factor $\epsilon_S/\epsilon_B$, where $\epsilon_S$ and $\epsilon_B$ are a signal and background efficiencies after applying the quark/gluon discrimination in multi-jet final states.  Therefore the ratio factor represents how many times $S/B$ increases after the application. 
We assume that all jets in signals are quark jets, i.e., $N_{q,S}=\Njets$, where $N_{q,S}$ is the number of quark jets in signals. 
Such signals will be also considered in the next section. 
In this assumption, the signal efficiency is naively estimated as $\epsilon_S \sim \epsilon_q^{N_{q,S}}$.
If a signal contains one more quark jet than a background, $S/B$ gains $\epsilon_q/\epsilon_g$ times by applying the quark/gluon tagging, i.e., $\epsilon_S/\epsilon_B=\epsilon_q/\epsilon_g$, where $\epsilon_q$ and $\epsilon_g$ are the quark and gluon jet efficiencies.
In case that the expected value of quark jets in background is $N_{q,B}$, we can expect that the improvement factor maximally increases up to $\epsilon_S/\epsilon_B \sim (\epsilon_q/\epsilon_g)^{N_{q,S}-N_{q,B}}$. 
Below, we estimate $N_{q,B}$ using generating functionals in Sec.~\ref{sec:app3}. 
Although the efficiency ratio $\epsilon_q/\epsilon_g$ is calculable for an IRC-safe and Sudakov safe variables, we calculate the ratio using a Monte-Carlo generator.

For the estimation of $N_{q,B}$, we first define the generating functional for proton collisions as,
\begin{align}
{\bf \Phi}_{pp\to {\rm jets}}
    \propto \sum_{i_1,i_2} f_{i_1}(x_{i_1},\mu_F) f_{i_2}(x_{i_2},\mu_F) {\bf \Phi}_{i_1i_2\to f_1f_2}
\end{align}
The starting scale, $\hat{p}_T$ and $p_{\rm beam}$ in ${\bf \Phi}_{i_1i_2\to f_1f_2}$ are set as in Sec.~\ref{Number_of_quark_jets}. 
The hard scale of collision is set to the invariant mass of initial partons $\sqrt{\hat{s}}$. 
We are interested in the case that the hard scale is a given new physics scale $\Lnew$, therefore we set as $\sqrt{\hat{s}}=\Lnew$. 
The transverse momentum of final states is set to half of the invariant mass as $\hat{p}_T=\sqrt{\hat{s}}/2=\Lnew/2$. 
The function $f_i$ is the proton PDF for an initial parton $i$, and the factorization scale is set to $\mu_F=\hat{p}_T$. The four-flavor scheme is used in the calculation. 
With these setting, we calculate the expected value of quark jets $N_{q,B}$ as in Sec.~\ref{Number_of_quark_jets}, 
\begin{align}
&N_{q,B} =
	\frac
	{\sum_{m=0}^n m \, R_{n,m}^{(pp\to {\rm jets})}}
	{R_{n}^{(pp\to {\rm jets})}}, \quad n = \Njets, \\
&R_{n}^{(pp\to {\rm jets})}  = 
	\left. \frac{1}{n!} \frac{\partial^n}{\partial u^n}
	\bold{\Phi}_{pp\to {\rm jets}} \right\vert_{u=0,v=1}, \\
&R_{n,m}^{(pp\to {\rm jets})}  = 
	\left. \frac{1}{n! m!} \frac{\partial^n}{\partial u^n} \frac{\partial^m}{\partial v^m} 
	\bold{\Phi}_{pp\to {\rm jets}} \right\vert_{u=v=0}. 
\end{align}

\begin{figure}[!t]
\begin{center}
\includegraphics[width=0.45\linewidth, bb=0 0 324 209]{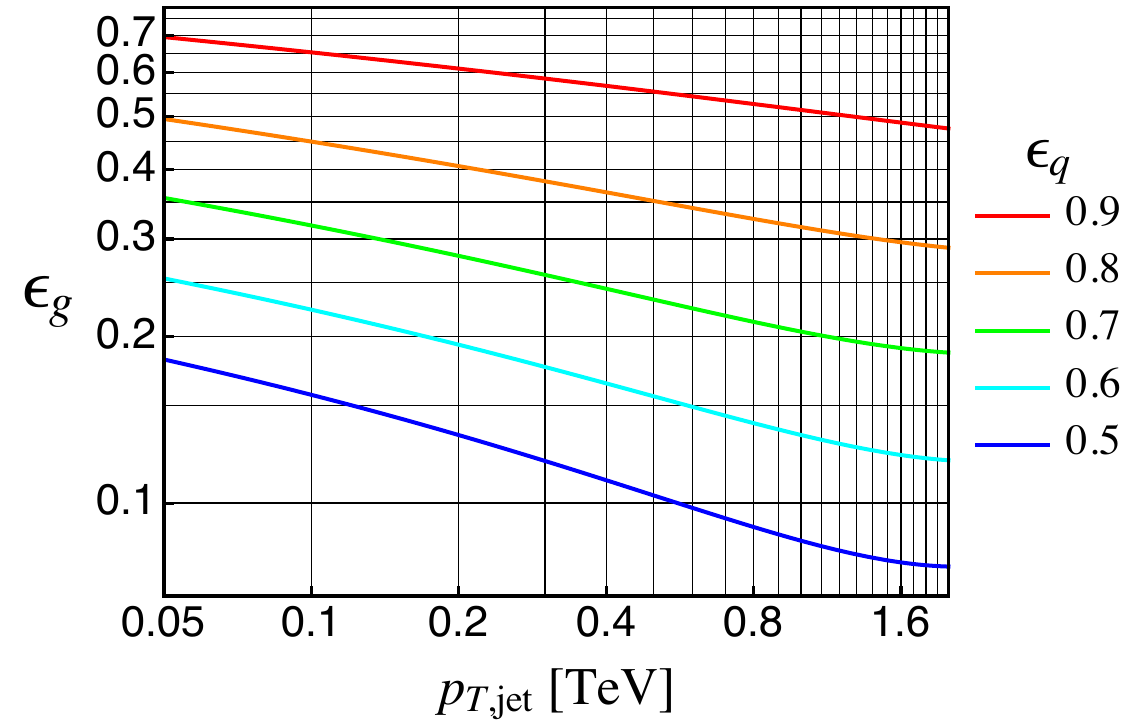}
\includegraphics[width=0.45\linewidth, bb=0 0 325 211]{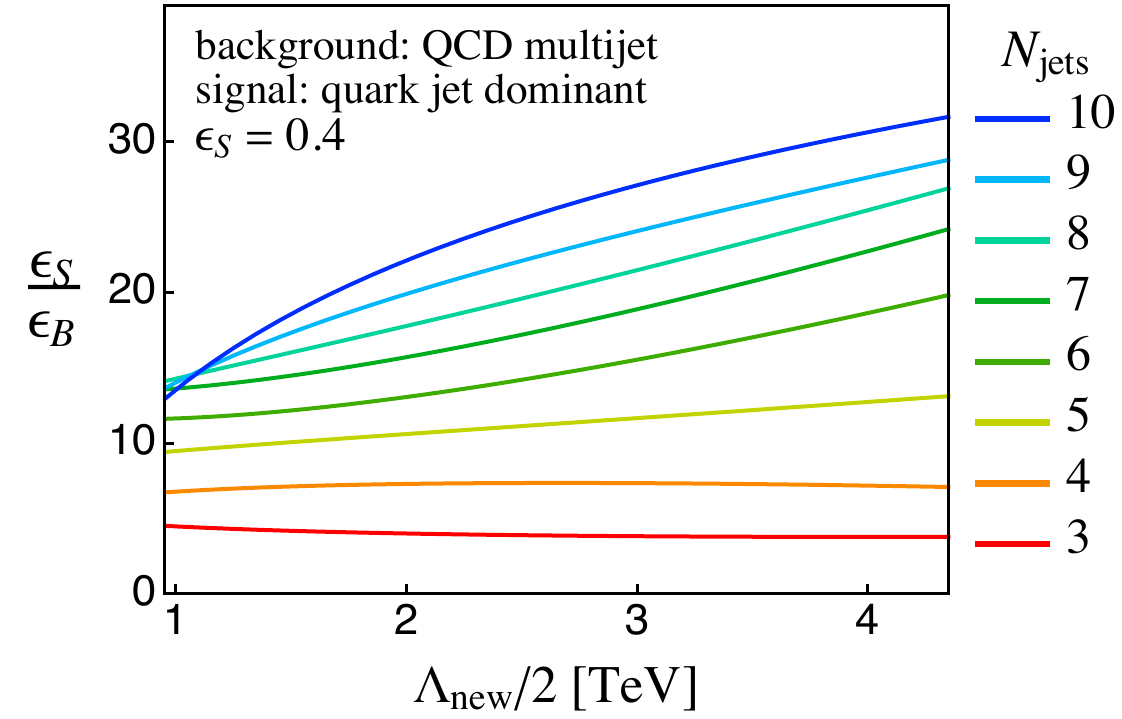}
\caption{{\footnotesize 
Jet $p_{T}$ dependence on gluon efficiencies for each quark efficiencies (left pannel). 
Dependence of a new physics scale $\Lnew$ on the improvement factor $\epsilon_S/\epsilon_B$ for each $\Njets$ categories (right pannel).
}}
\label{fig:SOB_analytic}
\end{center}
\end{figure}

Next, we show Monte-Carlo results for the quark and gluon jet efficiencies. 
In Fig.~\ref{fig:SOB_analytic}, the left figure shows the jet-$p_{T}$ dependence on the gluon efficiency for several quark efficiencies. 
We use {\tt Vincia} \cite{
Giele:2007di, Giele:2011cb, GehrmannDeRidder:2011dm, Ritzmann:2012ca, Hartgring:2013jma, Fischer:2016vfv} in the calculation.\footnote{
The one problem on the topic of quark/gluon discrimination is Monte-Carlo uncertainties of the predictions, and known that experimental data on certain observables related to quark/gluon tagging lie in between the predictions of the two MC generators {\tt Pythia} and {\tt Herwig} \cite{ATLAS:2012am,Khachatryan:2014hpa,Aad:2014gea}.  Although {\tt Vincia}'s results are close to {\tt Pythia}'s one, those lie in the the predictions of the two MC generators, and the uncertainties are focused in Refs.~\cite{Gras:2017jty, Sakaki:2015iya, Reichelt:2017hts}.
}  
We use an output evaluated by the BDT algorithm as a discrimination variable.  The output is trained using four variables, namely the number of charged tracks, energy correlation functions \cite{Larkoski:2013eya} with $\beta=0.2$ and $1.0$, and $p_T$-normalized jet mass ($m_{\rm jet}/p_T$).
Only charged track informations are used for the calculation of the BDT inputs. 
When we calculate $\epsilon_q$ and $\epsilon_g$ in the improvement factor, we set the jet $p_T$ in the estimation of the quark and gluon jet efficiencies to $\Lnew/\Njets$.

Finally, we estimate the improvement factor $\epsilon_S/\epsilon_B = (\epsilon_q/\epsilon_g)^{N_{q,S}-N_{q,B}}$ with the above setting. 
The right plot in Fig.~\ref{fig:SOB_analytic} shows the dependence of a new physics scale $\Lnew$ on the factor for each $\Njets$ categories.
The signal efficiency is fixed at 0.4.
We can see that the improvement factor increases as the number of jets increases since the difference of the number of quark jets between the signal and background, namely $N_{q,S}-N_{q,B}$, gets larger. 
Also, the factor improves as the new physics scale gets larger because discrimination power for the quark/gluon separation increases as the jet $p_T$ increases. 
The effect is clear in large $\Njets$ categories.
The probability that valence quarks flow into final states becomes larger as $\Lnew$ increases.
This makes the difference of the number of quark jets between the signal and background small, therefore the factor decreases. 
This effect is noticeable in small $\Njets$ categories.

\section{BSM searches in milti-jet final states}
\label{sec:BSM}
In this section, we calculate the improvement factor estimated in subsection 3.2 by a realistic data-driven way using a Monte-Carlo generator. 
The data-driven method is often used for the analysis of multi-jet final states. 
A typical analysis is the micro black hole search \cite{
Aad:2012hm, Chatrchyan:2012lia, CMS:2014exa, Aad:2016jxj, Khachatryan:2016xim, Sirunyan:2017cwe, Aaboud:2017faq}. 
In the analysis, phase space is divided by a variable related to the hard scale, e.g., the scalar sum of jet transverse momenta $H_T$, the scalar sum of the masses of large-$R$ jets, $M_J^{\sum}$ \cite{Hook:2012fd,Hedri:2013pvl,Cohen:2014epa} etc.
We fit the distribution of the variable on phase space at low-energy scale referred to as control region (CR), and estimate the number of background on phase space at high-energy scale referred to as signal region (SR) using the fit function. 
If there is excess from the estimated background, we think it as a sign of new physics.

One of the problems for such analyses is that we simplify the high jet multiplicity events too much. 
In the analysis explained above, only one or two inclusive variables are mainly used.\footnote{The sum of fat jet masses $M_J^{\sum}$ also contain some information on exclusive variables like jet $p_T$ and the distance between sub-jets.}
We can also utilize robust jet substructure variables for quark/gluon tagging in the analysis. 
It is difficult to predict multi-dimensional distributions for the jet substructure variables in multi-jet final states precisely, therefore, the data-driven approach is preferred to incorporate the jet flavor information in the analysis. 
We introduce a variable containing the information for data-driven analysis,\footnote{We found that the performance of the discrimination between the signal and background discussed below can increase slightly with a more complicated definition of $d$.   The optimization of variable $d$ is beyond the scope of this paper.}
\begin{align}
d = \sqrt{ \frac{1}{n} \sum_{i=1}^{n} Q_i^2 }, \quad \text{for $\Njets \geq n$,}
\end{align}
where $Q_i(>0)$ shows a kind of {\it quark-jet-ness} for $i$-th jet. 
If jet substructure for the $i$-th jet looks like quark jet rather than gluon jet, $Q_i$ takes larger value. 
In this paper, we use the BDT output used in Section~\ref{sec:Expected_improvement} as $Q_i$, which is trained like that quark and gluon jet are assigned to 1 and 0.  
The variable $d$ takes a large value for events which contain many quark jets.

We consider the following toy-signal topologies:
\begin{align}
(gg \text{ or } u\bar{u}) \to XX, \quad X \to n_X \text{-quarks}.
\end{align}
The pare production of a hypothetical heavy resonance $X$ has initial states $gg$ or $u\bar{u}$ in proton collisions and $X$ decays into $n_X$ quarks.
For example, the pair production of gluinos and squarks in SUSY with R-Parity Violation has the same decay topology.  
We generate hard processes using {\tt Madgraph5} \cite{Alwall:2014hca} with CTEQ6L PDF and have $X$ decay in phase space flatly.
When $n_X$ is odd or even, $X$ is assigned to the color-octet or -triplet respectively, and color indices of $X$ are connected to those of quarks in the large-$N_c$ limit.
We use {\tt Vincia} for the parton showering and the hadronization.

For the simulation of QCD multi-jet background, we use {\tt Vincia} with the default setting.

The generated signal and background are clustered with the anti-$k_t$ algorithm and the jet radius is set to $R=0.4$.  As selection cuts, minimum transverse momentum ($p_T>50{\rm GeV}$) and rapidity cut ($|\eta|<2.8$) are imposed to all jets.  The invariant mass of the collision system is set to $\sqrt{s}=14$ TeV.

\begin{figure}[!t]
\begin{center}
\includegraphics[width=0.95\linewidth, bb=0 0 911 434]{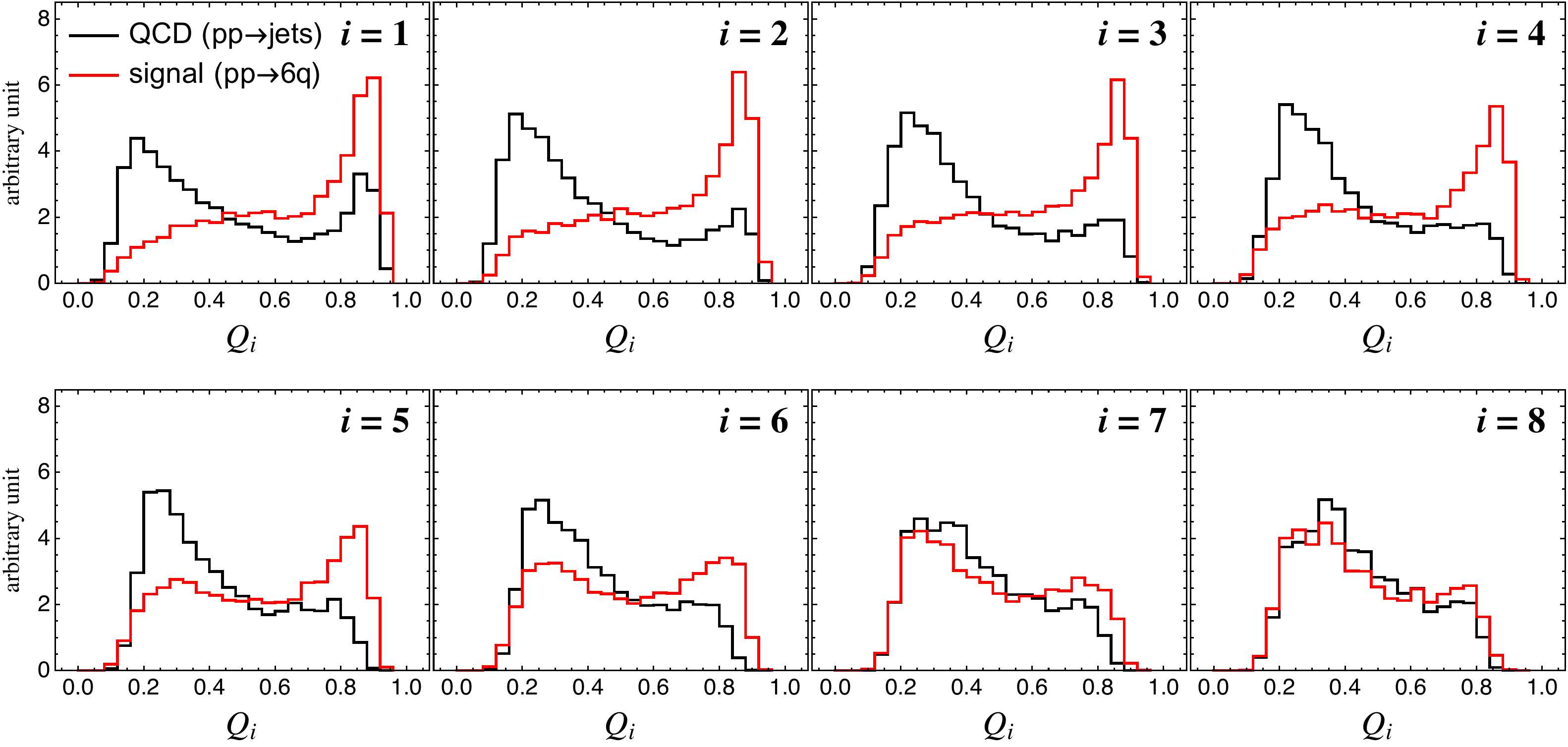}
\caption{{\footnotesize 
The black and red curves show the distribution of $Q_i$ for the background and signal, where $H_T>2$ TeV, $\Njets \geq 8$ are imposed. 
We use BDT outputs used in Section~\ref{sec:Expected_improvement} as $Q_i$, which is trained like that quark and gluon jet sample are assigned to 1 and 0.  
}}
\label{fig:bdt}
\end{center}
\end{figure}

In Fig.~\ref{fig:bdt}, the black and red curves show the distribution of $Q_i$ for the background and signal, where $H_T>2$ TeV, $\Njets \geq 8$ are imposed. 
For background, $Q_i$'s tend to be distributed in the region close to 0 since the gluon jets are dominant in QCD multi-jet final states, however, $Q_1$ has a clear peak on the side of 1 due to the effect from valence quark jets.  
For the signal, we set the mass of $X$ $(M_X)$ to 2 TeV, and $n_X=3$. 
The signal has six quarks in the hard process, so $Q_i$'s are distributed in the region close to 1 up to 6-th jet.  The 7-th and 8-th jets would stem from QCD radiations, so the differences between the signal and background become small.

\begin{figure}[!t]
\begin{center}
\includegraphics[width=0.35\linewidth, bb=0 0 300 243]{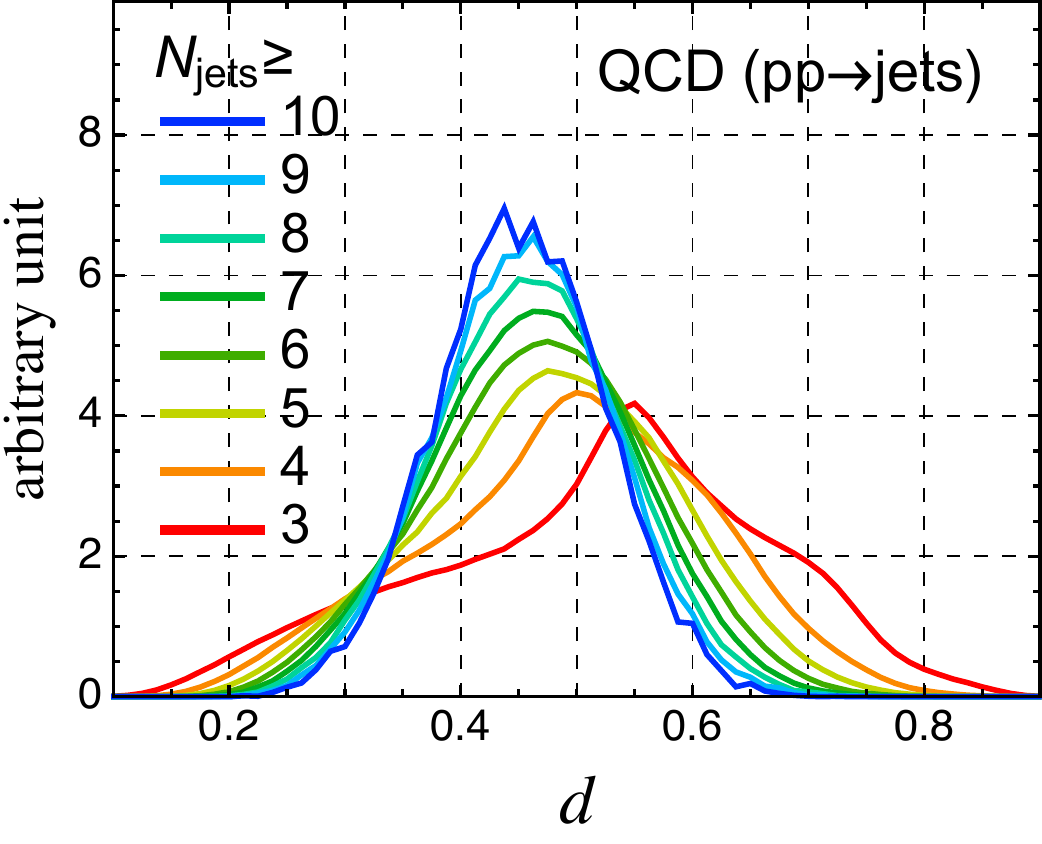}\hspace{20pt}
\includegraphics[width=0.35\linewidth, bb=0 0 300 243]{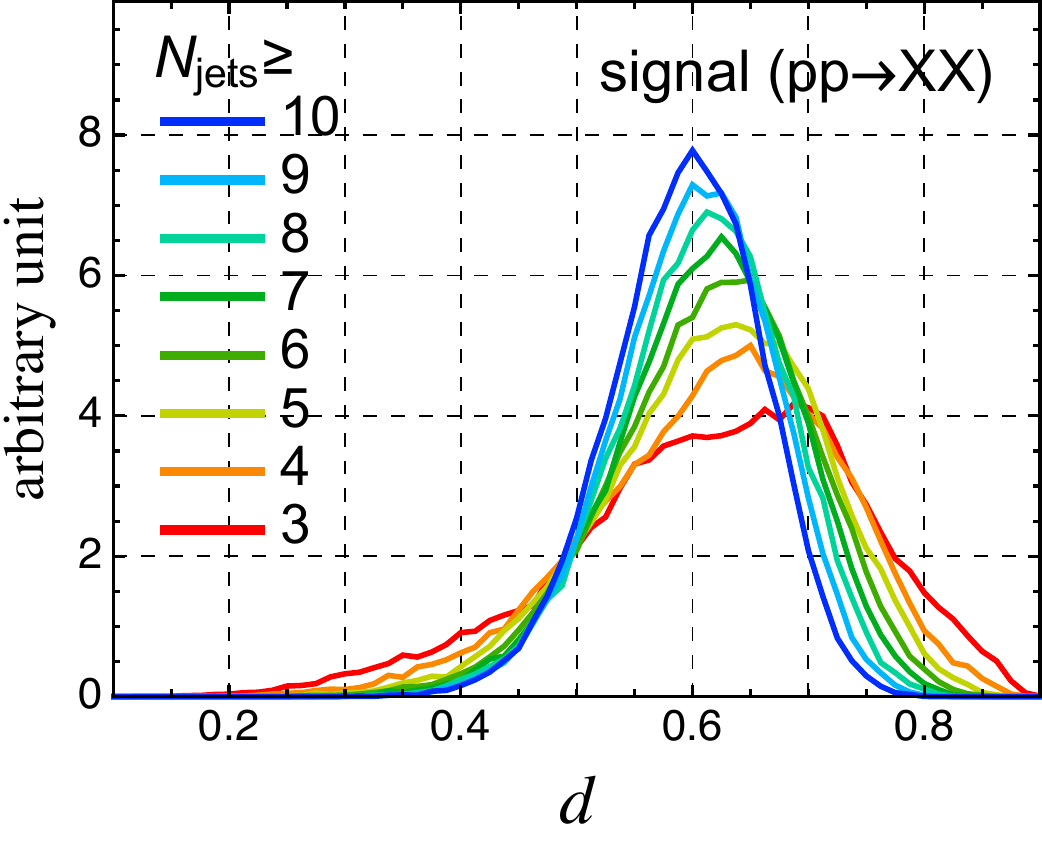}
\caption{{\footnotesize 
The distribution of $d$ for the QCD multi-jets (left) and the signal (right) for each $\Njets$ categories. 
The signal is set to $M_X=2$ TeV and $n_X=5$.
}}
\label{fig:d}
\end{center}
\end{figure}

In Fig.~\ref{fig:d}, the distribution of $d$ for the QCD multi-jets (left) and the signal (right) for each $\Njets$ categories are shown. 
The signal is set to $M_X=2$ TeV and $n_X=5$.
We can remove the background by imposing cut $d>d_{\rm cut}$, since the signal is distributed in large-$d$ region.

\begin{figure}[!t]
\begin{center}
\includegraphics[width=0.35\linewidth, bb=0 0 350 275]{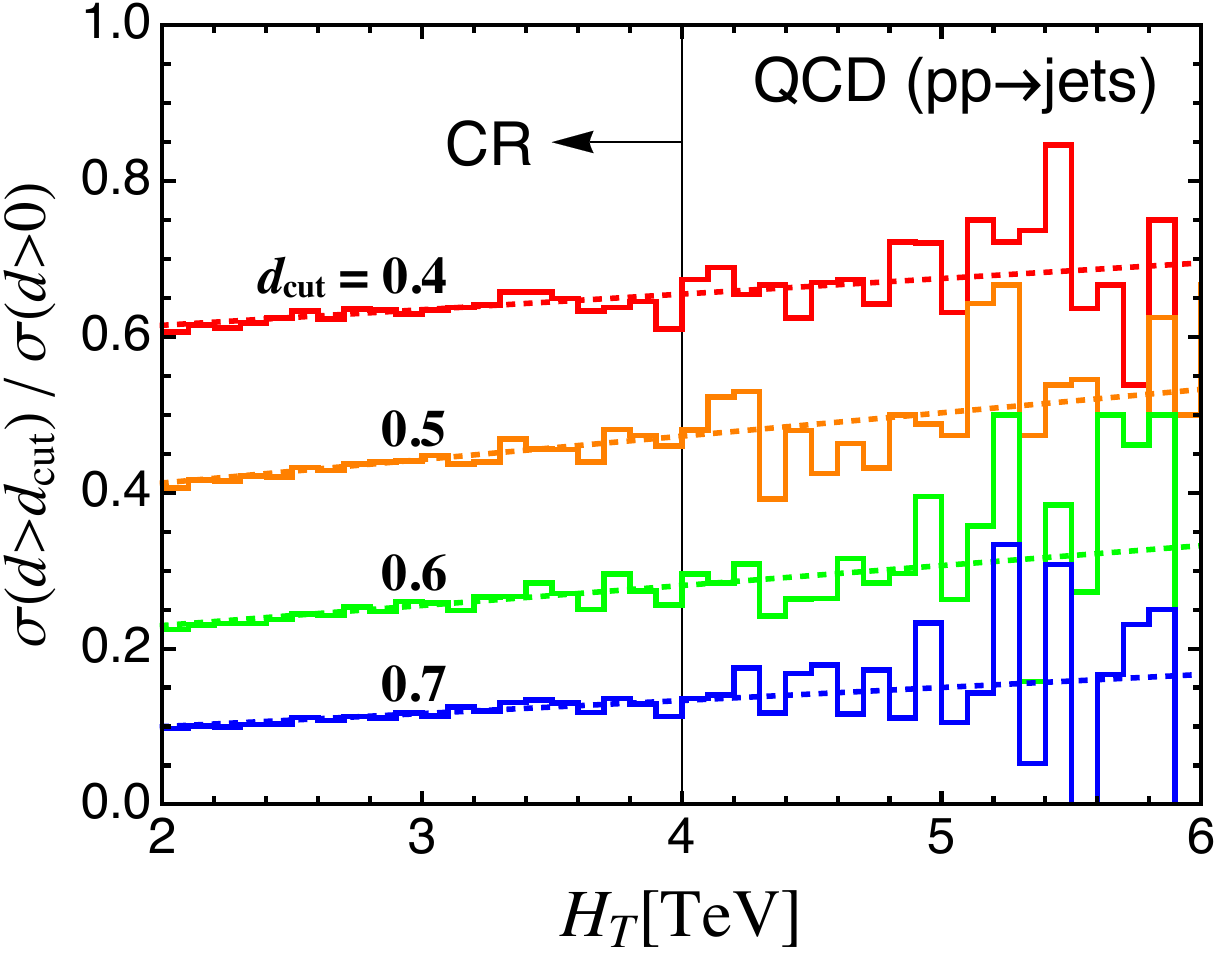}\hspace{20pt}
\includegraphics[width=0.35\linewidth, bb=0 0 350 275]{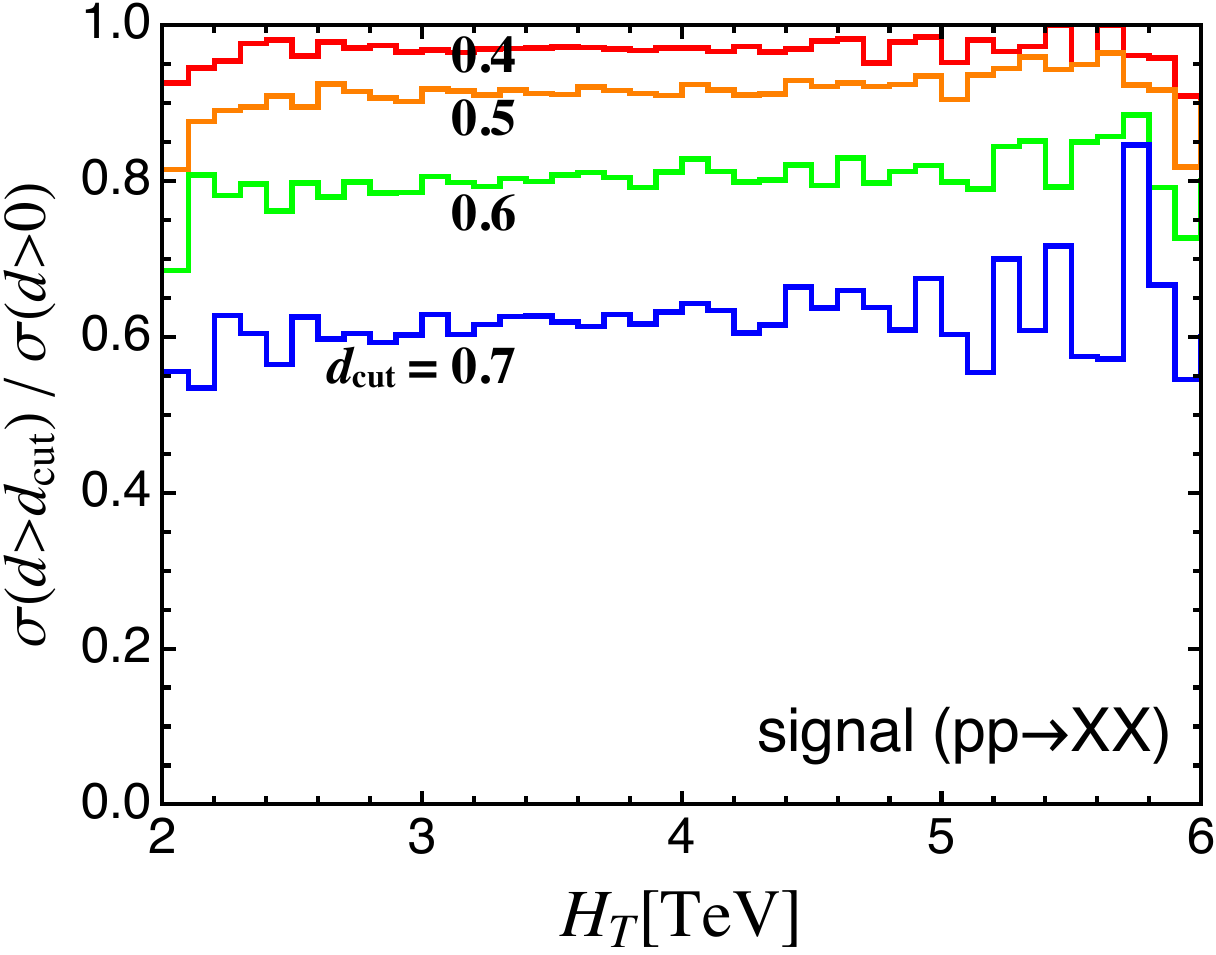}
\caption{{\footnotesize 
The remaining rates of the number of events after imposing $d$-cut for each $H_T$ bins.  The left and right figures are the results for the QCD multi-jets and the signal at $\Njets \geq 6$.  The signal parameters are $M_X=2$ TeV and $n_X=5$. 
The dotted curves show an example of interpolation curves which are fitted by using data in a control region, i.e., $H_T<4$ TeV.
}}
\label{fig:dcut}
\end{center}
\end{figure}

In Fig.~\ref{fig:dcut}, the remaining rates of the number of events after imposing $d$-cut for each $H_T$ bins are shown.  The left and right figures are the results for the QCD multi-jets and the signal at $\Njets \geq 6$.  The signal parameters are $M_X=2$ TeV and $n_X=5$. 
The background gets decreasing rapidly after imposing larger $d_{\rm cut}$. 
What we want to know is that the number of backgrounds after imposing $d$-cut at high energy region. 
Due to the complexity of large jet multiplicity events, such number should be estimated by the data-driven method. 
The dotted curves in Fig.~\ref{fig:dcut} show an example of interpolation curves which are fitted by using data in a control region, i.e., $H_T<4$ TeV in the figure.
In the practical analysis, we can know the ratio of background in signal regions using such interpolation curves, and can obtain the upper bound on the cross-section imposed $d$-cut in QCD multi-jet final states.

We estimate how much $S/B$ improves by applying the quark/gluon discrimination in multi-jet final states. 
The ratio is given by,
\begin{align}
&\frac{S}{B} \simeq \frac
{\sigma_S(\text{selection cuts} ~\&~ H_T \text{-cut})}
{\sigma_B(\text{selection cuts} ~\&~ H_T \text{-cut})}
\times
\frac{\epsilon_S(d_{\rm cut})}{\epsilon_B(d_{\rm cut})}, \label{SOB1} \\
&\epsilon_X(d_{\rm cut}) = 
\frac
{\sigma_X(\text{selection cuts} ~\&~ H_T \text{-cut} ~\&~ d>d_{\rm cut})}
{\sigma_X(\text{selection cuts} ~\&~ H_T \text{-cut} ~\&~ d>0) \hspace{12pt}},
\end{align}
where $\sigma_X$ is the cross-section for the signal $(X=S)$ and the background $(X=B)$ after imposing the condition in the brackets. 
The impact coming from quark/gluon discrimination is included in the second ratio $\epsilon_S(d_{\rm cut}) / \epsilon_B(d_{\rm cut})$.
We employ a $H_T$-cut $H_T>1.8M_X$, which almost makes the significance of signal maximum in the case that systematic uncertainties and $d$-cut are neglected.

\begin{figure}[!t]
\begin{center}
\includegraphics[width=0.98\linewidth, bb=0 0 1030 214]{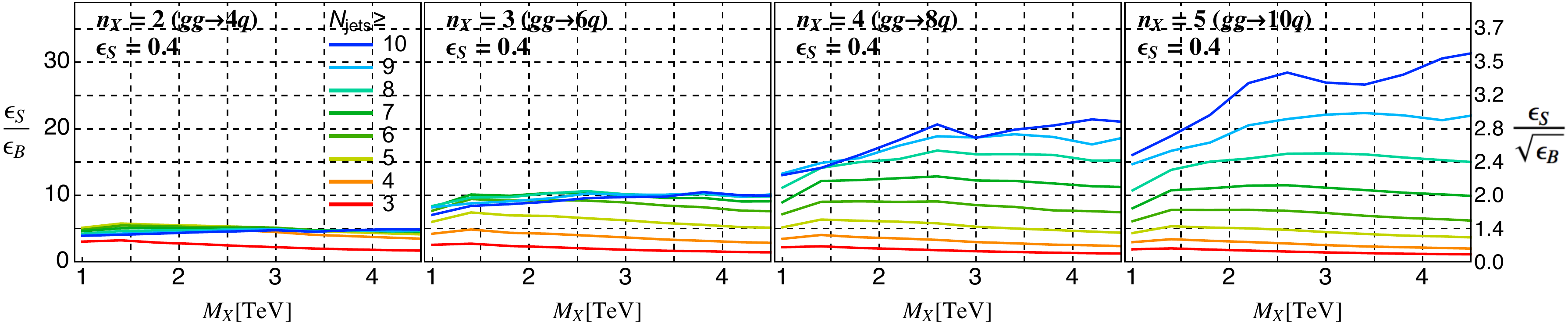}
\caption{{\footnotesize 
$M_X$-dependence on the efficiency ratio. 
We can see how the ratio changes with increasing the lower bound of $\Njets$ from 3 to 10, and $n_X$ from 2 (left-most) to 5 (right-most). 
}}
\label{fig:SOB_gg}
\end{center}
\end{figure}

In Fig.~\ref{fig:SOB_gg}, $M_X$-dependence on the efficiency ratio is shown, where the dependence on $\epsilon_S/\sqrt{\epsilon_B}$ is also shown on the right axis.
We can see how the ratio changes with increasing the lower bound of $\Njets$ from 3 to 10, and $n_X$ from 2 (left-most) to 5 (right-most). 
We choose $d_{\rm cut}$ which gives the signal efficiency $\epsilon_S=0.4$. 
These are the results in the case that the initial state is $gg$. 
The ratio clearly keep increasing until the lower bound of $\Njets$ reaches up to $2n_X$ since $2n_X$ quarks are contained in the hard processes of signal. 
A quark emitted from $X$ could be softer than partons arising from initial and/or final state radiations.  
In that case, the quark from $X$ could be $(2n_X+1)$-th jet, so we can see some improvements on $S/B$ even if the lower bound of $\Njets$ is greater than $2n_X$. 
We can understand the behavior of $M_X$ dependence on the improvement factor from the result in Section \ref{sec:Expected_improvement}.
The ratio improves as $M_X$ get larger because the discrimination power of the quark/gluon separation increases as the jet $p_T$ increases. 
The effect is clear in the large $\Njets$ categories.
The probability that valence quarks are in final states becomes larger as the masses increase.  This makes the difference of the number of quark jets between the signal and background small, and makes the ratio decrease. 
The effect looks important in the small $\Njets$ categories. 
We also see a good agreement between the right-most figure in Fig.~\ref{fig:SOB_gg} and the semi-analytic result in Fig.~\ref{fig:SOB_analytic}.  In both cases, signals are quite quark jet dominant.  
Note that the hard scale or the invariant mass for the pair production $X$ is about $2M_X$, therefore, the label $\Lnew/2$ on the $x$-axis in Fig.~\ref{fig:SOB_analytic} almost corresponds to $M_X$ on the $x$-axis in Fig.~\ref{fig:SOB_gg}.

\begin{figure}[!t]
\begin{center}
\includegraphics[width=0.98\linewidth, bb=0 0 1030 214]{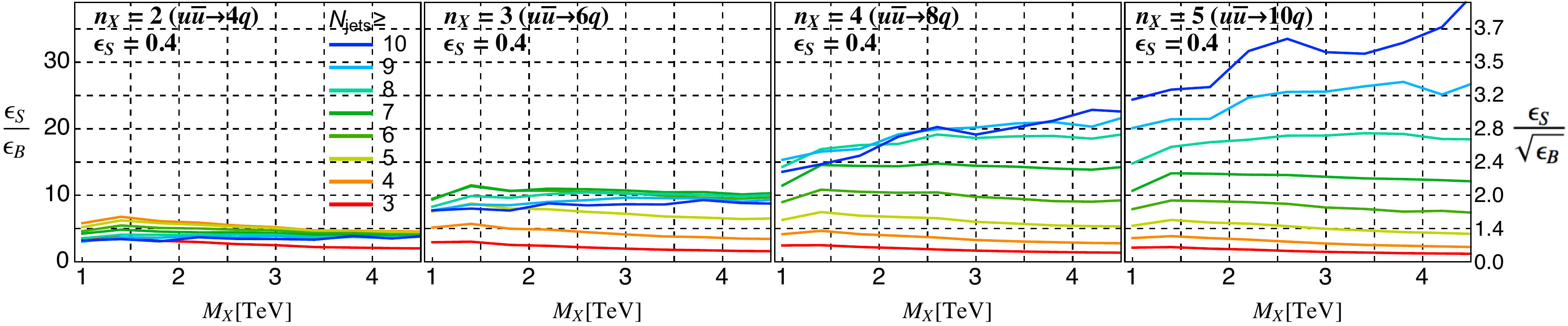}
\caption{{\footnotesize 
Same as Fig.~\ref{fig:SOB_gg}, with the initial states $u\bar{u}$. 
}}
\label{fig:SOB_uu}
\end{center}
\end{figure}

Fig.~\ref{fig:SOB_uu} shows the same as Fig.~\ref{fig:SOB_gg}, with the initial states $u\bar{u}$. 
The Born configuration of $gg$ initial state tends to emit valence quark jets at high energy, but also emit gluon jets more compared to the case of $u\bar{u}$ initial state since the color factor for $g\to gg$ is larger than $q\to qg$ 9/4 times. 
Consequently, initial state radiations from $gg$ reduce the quark jet fraction, then the improvement factors for $u\bar{u}$ is slightly better than those for $gg$.

\section{Conclusions}
\label{Conclusions}

The quark/gluon discrimination is maximally utilized for searches of new physics that predicts quark and gluon jet fractions which is different from what the QCD background does. 
To know the jet flavor structure in QCD multi-jet final states at hadron colliders,
we have introduced quark jet rates $R_{n,m}$ which is the probability that a parton or a matrix element produce $n$ jets in which $m$ quark jets are contained.  
We have calculated generating functionals, which contain the quark jet rates as coefficients, for initial and final state by using the QCD resummation technique.

Exponential structures of the functionals are evaluated and we can get the quark jet rates $R_{n,m}$ from the expansion coefficients.
The increment of gluon jets mainly arises from leading logarithmic terms in the coefficients, and that of quark jets comes from next-to-leading logarithmic terms.
More details of the logarithmic structure are also shown. 
In order to know a rate of the increment of quark jets, 
we have shown the expected value of the number of quark jets in $\Njets$ categories for matrix element configurations $gg\to gg$, $gu\to gu$ and $uu\to uu$.
For example, when we set the jet radius, jet $p_T$ cut, and the parton $p_T$ cut to $R=0.4$, $p_0=50$ TeV and $\hat{p}_T=1$ TeV, the number of quark jets increases by about 0.25, 0.18 and 0.12 for the three configurations while the number of jet increases by 1. 
We have also checked the consistency between the analytical results and Monte-Carlo predictions.

Since the QCD multi-jets are basically composed of few valence quark jets and many gluon jets, we expect a big improvement on $S/B$ for a signal which predicts many quark jets, by using the quark/gluon discrimination. 
We have estimated the improvement semi-analytically using the above results and have shown that the improvement gets larger as the number of quark jets in signals increases. 
For example, $S/B$ increases by about 20 times in the case that a new physics scale is $\Lnew=4$ TeV and the number of quark jets is 10.

We have introduced a variable $d$ that takes a large value for events in which many quark jets are contained, and have suggested a data-driven analysis using the variable.
Assuming a pare production of a hypothetical heavy resonance $X$ which decays into $n_X$ quarks as a signal, we have evaluated the large improvement on $S/B$ for each masses of $X$, $n_X$ and initial states in Monte-Carlo analysis, and have shown the usability of quark/gluon discrimination in multi-jet final states.

\section*{Acknowledgements}
I thank Mihoko M. Nojiri, Hideki Okawa, Bryan R. Webber and Jesse Thaler for helpful discussions and useful comments.  This work is supported by Samsung Science and Technology Foundation under Project Number SSTF-BA1602-04.

\appendix

\section{Details on quark jet rates}

\subsection{Formulae}
\label{app:formulae}

In Sec.~\ref{sec:app1}, generating functionals which contain effects of emissions coming from only progenitor partons are evaluated, and those for final states are given as,
\begin{align}
\Phi_q^{(\LL+\qq)} &= uv \, \Delta_q^{1-u}, \\
\Phi_g^{(\LL+\qq)} &= u   \, \Delta_g^{1-u} \Delta_{\qq}^{1-uv^2}. 
\end{align}
For initial states, we get
\begin{align}
\Psi_i^{(\LL+\qq)} &= \Pi_{i,1}^{1-u}   \Pi_{i,2}^{1-uv},  \quad  i\in \{q,g\}.
\end{align}

In Sec.~\ref{sec:app2}, we have also considered effects of subsequent emissions. 
The functionals are factorized into the primary terms and exponential terms related to the effect as,
\begin{align}
\Phi_q^{(\LL+\qq+\sub)} &= \Phi_q^{(\LL+\qq)} \times  \exp(S_q), \\
\Phi_g^{(\LL+\qq+\sub)} &= \Phi_g^{(\LL+\qq)} \times  \exp(S_g)  \exp(S'),
\end{align}
where
\begin{align}
& S_i = -u       \ln\De_{i}     \, I_1(p\ka\la, q\la), \\
& S'   = -uv^2 \ln\De_{\qq} \, I_2(w), 
\end{align}
and 
\begin{align}
& I_1(x,y) = \frac{\Ein(x+y)-\Ein(y)}{x}-1, \\
& I_2(z) = \frac{e^z-1}{z}-1, \\
& p = (1-u)a_g, \quad q \simeq (2/3)(1-uv^2)a_{\qq}, \label{eq:p_q}\\
& \Ein(z) = \sum_{n=1}^{\infty} \frac{-(-z)^n}{n \, n!}.
\end{align}
For $q$ in Eq.~(\ref{eq:p_q}), we used an approximation $c_{\qq}\simeq 2/3$.
For initial states, the generating functionals containing effects of subsequent emissions are
\begin{align}
\Psi_i^{(\LL+\qq+\sub)} &= \Psi_i^{ (\LL+\qq)} \times  \exp( S_i[f_{i/i}] )  \exp(S_i'), 
\end{align}
where 
\begin{align}
& S'_q =-uv \ln\Pi_{q,2}  I_2 (- \overline{w}), \\
& S'_g =-uv \ln\Pi_{g,2}  I_2 (+\overline{w}). 
\end{align}
You can see the definition of $\overline{w}$ in Eq.~(\ref{eq:wbar}).

In Sec.~\ref{sec:app3}, the running effects of $\alpha_s$ are also considered.  In the case, the functionals for final states are written as,
\begin{align}
\Phi_q^{(\LL+\qq+\sub+\delta\alpha_s)} &= \Phi_q^{(\LL+\qq+\sub)} \times \exp(\tilde{S}_q) \exp(T_q), \\
\Phi_g^{(\LL+\qq+\sub+\delta\alpha_s)} &= \Phi_g^{(\LL+\qq+\sub)} \times \exp(\tilde{S}_g) \exp(T_g) \exp(\tilde{S}') \exp(T'),
\end{align}
where 
\begin{align}
& \tilde{S}_i = u \ln\De_i \left(-\frac{\ka+\la}{2} + I_3(s) - I_3(q) \right), \\
& T_i   = -(1-u)\ln\De_i \, I_4(a\ka,a\la), \\
& \tilde{S}'  \simeq -uv^2 a_{\qq} \frac{a}{1+a\ka+a\la}
	\left[ \left( \ka - \frac{13}{12} \right)\left( \frac{e^w-1}{w}-1 \right) + \la \left( \frac{1-e^w}{w^2} + \frac{e^w}{w} - \frac{1}{2} \right) \right] , \label{eq:tildeSdash}\\
& T'     \simeq (1-uv^2)  \frac{2}{3}a_{\qq}\la  \frac{a}{1+a\ka+a\la} \left(\ka+\frac{\la}{2}-\frac{13}{12} \right), \label{eq:Tdash}
\end{align}
and
\begin{align}
& I_3(z) = -\frac{1}{p\ka\la} \left[  \left( \frac{z}{p} + \la \right)I_2(-z\la) + \frac{q}{p}\Ein(z\la)  \right], \\
& I_4(x,y) = 1 - \frac{(1+x+y)\ln(1+x+y) - (1+x)\ln(1+x) - (1+y)\ln(1+y)}{xy}.
\end{align}
In the calculations of $\tilde{S}'$ and $T'$ in Eqs.~(\ref{eq:tildeSdash}) and (\ref{eq:Tdash}), we use the following approximation,
\begin{align}
\int_0^{\ka} d\ka' \int_0^{\la} d\la' F(\ka',\la') D
 = 
\int_0^{\ka} d\ka' \int_0^{\la} d\la' F(\ka',\la') \frac{a(\ka'+\la')}{1+a(\ka+\la)}.
\end{align}
This approximation has good accuracy because the integrants for $\tilde{S}'$ and $T'$ are localized around $\ka'=\ka$ and $\la'=\la$.  
For initial states, we get
\begin{align}
\Psi_i^{(\LL+\qq+\sub+\delta\alpha_s)} 
	&= \Psi_i^{(\LL+\qq+\sub)}  \times  \exp(\tilde{S}_i[f_{i/i}])  \exp(T_i[f_{i/i}])   \exp(\tilde{S}_i')  \exp(T_i'), 
\end{align}
where
\begin{align}
& \tilde{S}'_q  \simeq uv \, \ln\Pi_{q,2} \, a I_5(c_{g/q}^{(1)}, c_{g/q}^{(2)}), \label{eq:tildeSdash_q}\\
& \tilde{S}'_g  \simeq uv \, \ln\Pi_{g,2} \, a I_5(c_{Q/g}^{(1)}, c_{Q/g}^{(2)}), \label{eq:tildeSdash_g}\\
& T'_q  = (1-uv) \, \frac{a_{\qq}}{n_f} \, \delta c_{g/q}^{(1)} \la, \\
& T'_g  = (1-uv) \, a_q                       \, \delta c_{Q/g}^{(1)} \la, 
\end{align}
and
\begin{align}
& I_5(c_1,c_2) = \left(  \frac{1-e^{\wB}+\wB e^{\wB}}{\wB^2} - \frac{1}{2}  \right) \la + \frac{e^{\wB}-1-\wB}{\wB} \cdot \frac{c_2}{c_1}, \\
&\delta c_{g/q}^{(1)}  =
\frac{n_f}{a_{\qq}} \int_0^{\kaB} d\kaB' \Gamma_{g\to\qq}(\kaB') \frac{f_g(x')}{f_q(x')} 
\left[  1-\frac{1}{a\la}\ln  \left(  1+\frac{a\la}{1+a\kaB'}  \right)  \right]
,\\
&\delta c_{Q/g}^{(1)} = 
\frac{1}{a_q} \int_0^{\kaB} d\kaB' \sum_q \Gamma_{q\to gq}(\kaB') \frac{f_q(x')}{f_g(x')} 
\left[  1-\frac{1}{a\la}\ln  \left(  1+\frac{a\la}{1+a\kaB'}  \right)  \right]
.
\end{align}
In the calculations of $\tilde{S}'_{q,g}$ in Eqs.~(\ref{eq:tildeSdash_q}) and (\ref{eq:tildeSdash_g}), we expand $D$ by the logarithms and take into account only the leading term, namely $D\simeq a(\ka'+\la')$.

\subsection{Matrix element corrections}
\label{app:MEC}

\begin{figure}[!ht]
\begin{center}
\includegraphics[width=0.5\linewidth, bb=0 0 300 233]{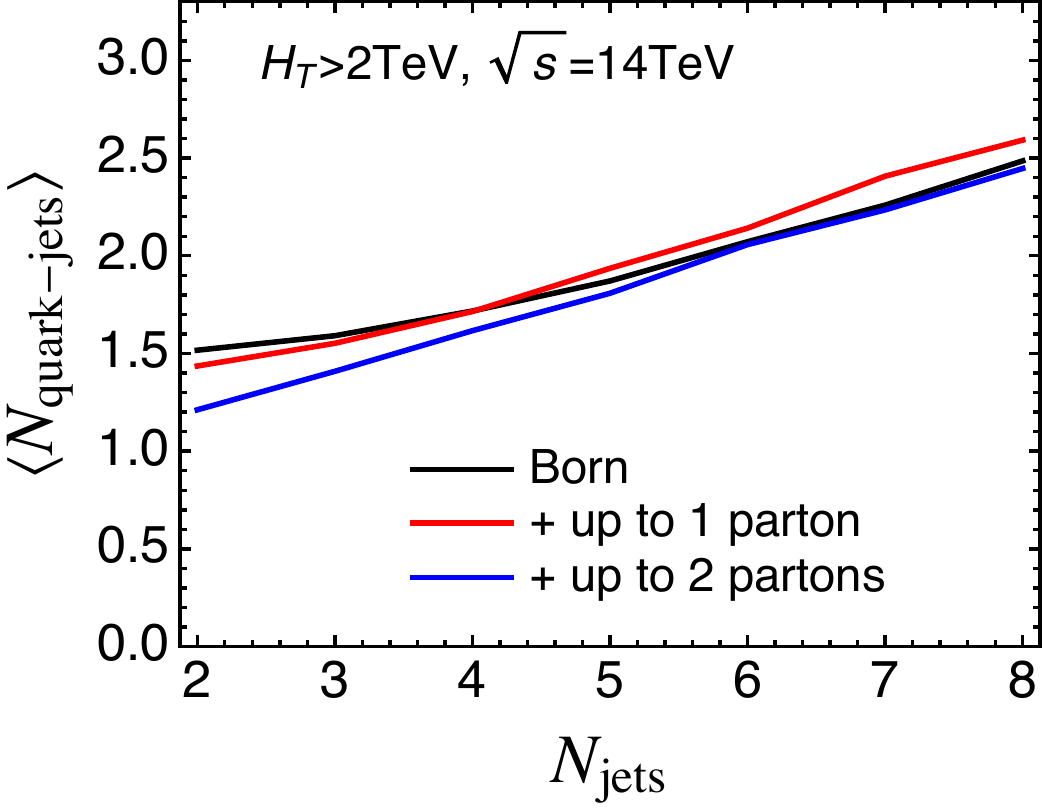}
\caption{{\footnotesize
Matrix element correction to the number of quark jets with CKKW matching. 
The black curve is a result for $pp\to jj$ + parton showers. 
One and two partons are matched into the Born and the results are shown in the red and blue curves, 
where we impose $H_T>2$ TeV and set to $\sqrt{s}=14$ TeV. 
}}
\label{fig:match}
\end{center}
\end{figure}

In Sec.~\ref{sec:Numerical-results}, the number of quark jets for each $\Njets$ categories are evaluated by applying parton showers to Born configurations.  In the calculation, matrix element corrections are absence for more than 2 jets. 
In Fig.~\ref{fig:match}, we show the matrix element correction to the number of quark jets with CKKW matching using {\tt Sherpa} \cite{Gleisberg:2003xi, Gleisberg:2008ta}.
The black curve is a result for $pp\to jj$ + parton showers. 
One and two partons are matched into the Born configuration and the results are shown in the red and blue curves. 
We impose $H_T>2$ TeV and set to $\sqrt{s}=14$ TeV.

In the black curve, $\langle N_{\text{quark-jets}} \rangle$ is about 1.5 at $\Njets =2$.  This means that final states tend to become two valence quarks. In this case, the curve has an artificial kink at $\Njets=3$ as discussed in Sec.~\ref{sec:Numerical-results}.  We can see that the kink disappears with matching. 
In the red and blue curves, the matrix element corrections are contained up to $\Njets=3$ and 4, and we find the configuration containing gluons in final states increases.
As we see in Sec.~\ref{sec:Numerical-results}, the increase rate of quark jets for the gluon final state is larger than that for the quark final state. 
Therefore, the rate slightly increases after the matching.

\subsection{Estimating the number of quark jets in Monte-Carlo samples}
\label{app:qjet_def}
In Fig.~\ref{fig:Nq_all} of Sec.~\ref{Number_of_quark_jets}, we compare the number of quark jets calculated analytically with that using Monte-Carlo events at parton level.
We define a jet flavor for each jet in multi-jet final states and estimate the expected value of the number of quark jets $\langle N_{\text{quark-jets}} \rangle$ for each $\Njets$ category. 
One can define a quark or gluon jet with jet constituents. 
First, we look for a $q\bar{q}$ pair in the constituents and convert the pair to a gluon, and add the gluon into the constituent list. 
We continue this until any pair cannot be found. 
After the conversion, if only gluons are in the list we call the jet as gluon jet otherwise quark jet. 
If a quark jet contains only one quark in the list after the conversion we call it as well-defined quark jet otherwise ill-defined quark jet. 
In the definition, the number of quark jets is IR-unsafe because the number can change due to a quark of $q\bar{q}$ that a soft gluon decays into. 
We temporarily call the effect from the soft gluons as $g_{\rm soft} \to q\bar{q}$ pollution. 
This pollution turns a gluon jet or a well-defined quark jet into a well-defined quark jet or an ill-defined quark jet.\footnote{We ignore that more than one quark contamination gets into a single jet and also that $g_{\rm soft} \to q\bar{q}$ pollution turns a well-defined quark jet into a gluon jet.}

In consideration of the effect, we reduce IR-unsafety by adding a correction to the number of quark jet calculated by the above simple algorithm.
We first define two probabilities $P_{n,m}$ and $\hat{P}_{n,m}$ that show quark jet rates calculated in a system where we can ignore $g_{\rm soft} \to q\bar{q}$ pollution and cannot. 
Here, the quark jet rate shows the probability that events contain m quark jets and totally $n$ jets. 
$\hat{P}_{n,m}$ can be expressed by the following sum,
\begin{align}
\hat{P}_{n,m} = \hat{P}_{n,m}^{\rm (well)} + \hat{P}_{n,m}^{\rm (ill)}.
\end{align}
$\hat{P}_{n,m}^{\rm (well)}$ shows the probability that events contain $m$ quark jets and totally $n$ jets, where the quark jets are all well-defined jets. 
$\hat{P}_{n,m}^{\rm (ill)}$ is the probability for the case that the ill-defined jet is contained. 
Using Monte-Carlo samples, we can calculate the two probabilities with the jet flavor definition explained above paragraph.

The ill-defined probability is approximately given as,
\begin{align}
\hat{P}_{n,m}^{\rm (ill)} \simeq m  \epsilon  P_{n,m}, 
\end{align}
where $\epsilon$ shows the probability that $g_{\rm soft} \to q\bar{q}$ pollution get into a jet. 
This shows a contribution that a well-defined quark jet turns into an ill-defined quark jet by the pollution. 
There is another relation, 
\begin{align}
\hat{P}_{n,m}^{\rm (well)} \simeq (1-n\epsilon) P_{n,m} + [n-(m-1)] P_{n,m-1}, 
\end{align}
The 1st term is the contribution for the case that $g_{\rm soft} \to q\bar{q}$ pollution doesn't affect to the jet flavor of any jets.  
The 2nd term is the contribution for the case that a gluon jet turns into well-defined quark jets by the pollution. 
Then, a ratio of the measurable probabilities is,
\begin{align}
\frac{\hat{P}_{n,m}^{\rm (ill)}}{\hat{P}_{n,m}} = m \epsilon + \mathcal{O}(\epsilon^2). \label{eq:epsilon_exp}
\end{align}
We denote the expected value of the number of quark jets in a system where $g_{\rm soft} \to q\bar{q}$ pollution can be neglected or cannot be neglected as $\langle N_{\text{quark-jets}} \rangle$ or $\langle \hat{N}_{\text{quark-jets}} \rangle$, and the values can be related to the quark jet rates as,
\begin{align}
\langle N_{\text{quark-jets}} \rangle = \frac{\sum_{m=0}^{n} m P_{n,m}}  {\sum_{m=0}^{n} P_{n,m}}, \quad
(N,P) \leftrightarrow (\hat{N},\hat{P}). 
\end{align}
Finally, we obtain the following expression,
\begin{align}
\langle N_{\text{quark-jets}} \rangle = \langle \hat{N}_{\text{quark-jets}} \rangle
   - (n-\langle \hat{N}_{\text{quark-jets}} \rangle) \epsilon + \mathcal{O}(\epsilon^2). \label{eq:NqMod}
\end{align}
Ignoring $O(\epsilon^2)$ term, we estimate $\epsilon$ from Eq.~(\ref{eq:epsilon_exp}) and correct the $\langle \hat{N}_{\text{quark-jets}} \rangle$ to $\langle N_{\text{quark-jets}} \rangle$ with Eq.~(\ref{eq:NqMod}). 
We employ $\langle N_{\text{quark-jets}} \rangle$ as the Monte-Carlo results, which is more appropriate to compare with analytical results in terms of IR-safety.

One may come up with a way to reduce IR-unsafety from $g_{\rm soft} \to q\bar{q}$ pollution with flavour-$k_t$ algorithm \cite{Banfi:2006hf}, but it's not simple to apply it on multi-jet final states.
In our analytical calculation jet is defined by jet-radius $R$ and $p_{T,{\rm cut}}$, and in order to cluster jet based on the definition, it is necessary to partially modify the algorithm. 
First, in order to cluster unresolved emission whose distance between a jet core is less than $R$, it is necessary to modify the distance factor in the measure of flavor-$k_t$ algorithm for hadron collider $d_{ij}^{(F)}$ as $\Delta\eta_{ij}^2+\Delta\phi_{ij}^2 \to (\Delta \eta_{ij}^2+\Delta\phi_{ij}^2)/ R$. 
Also, the beam measure $d_{iB}^{(F)}$ for quark (or {\it flavoured} particle) in this algorithm is larger than jet-$p_T$ employed in commonly used algorithms. 
As a result, when a jet is clustered with the inclusive variant of the algorithm without introducing $d_{\rm cut}$, hard particles separated by $R$ or more can be clustered. 
This doesn't match our jet definition. 
If we introduce $d_{\rm cut}$ and cluster jets with the exclusive variant, it would be appropriate to choose as $d_{\rm cut}=p_{T,{\rm cut}}$.
If $d_{\rm cut}$ is set to greater or less than $p_{T, {\rm cut}}$, the minimum value of jet $p_T$ will be greater or less than $p_{T,{\rm cut}}$. 
However, even if we set as $d_{\rm cut}=p_{T,{\rm cut}}$, $d_{ij}^{(F)}$ and $d_{iB}^{(F)}$ become easily larger than $d_{\rm cut}$ as increasing the hard process scale. 
In the case that $d_{ij}^{(F)}$ and $d_{iB}^{(F)}$ are larger than $d_{\rm cut}$ for $i$ and $j$, the two partons are declared as jets even if the distance between $i$ and $j$ is less than $R$. 
This also doesn't match our jet definition. 
While avoiding problems such as above, we might be able to devise a good way to define IR-safe inclusive jets in multi-jet final states, but this time we counted the number of quark jets in the way written in the above paragraph.

\subsection{Initial and final state radiation}
\label{app:ISR_FSR}

\begin{figure}[!ht]
\begin{center}
\includegraphics[width=0.32\linewidth, bb=0 0 300 230]{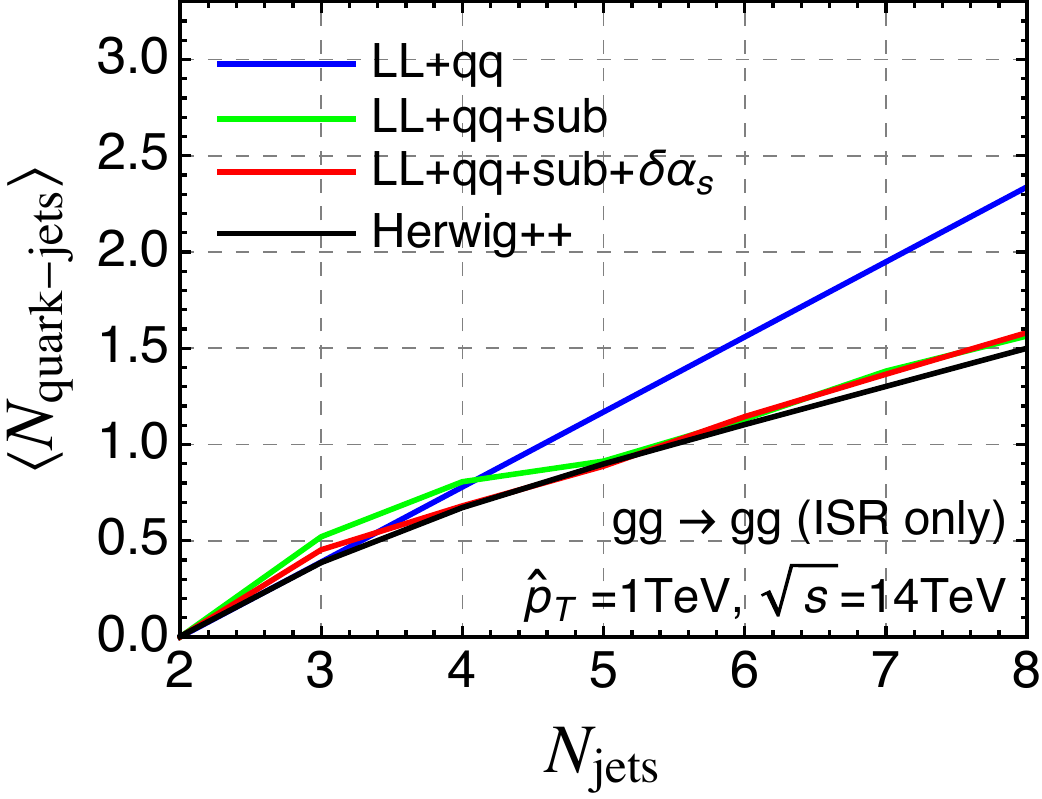}
\includegraphics[width=0.32\linewidth, bb=0 0 300 230]{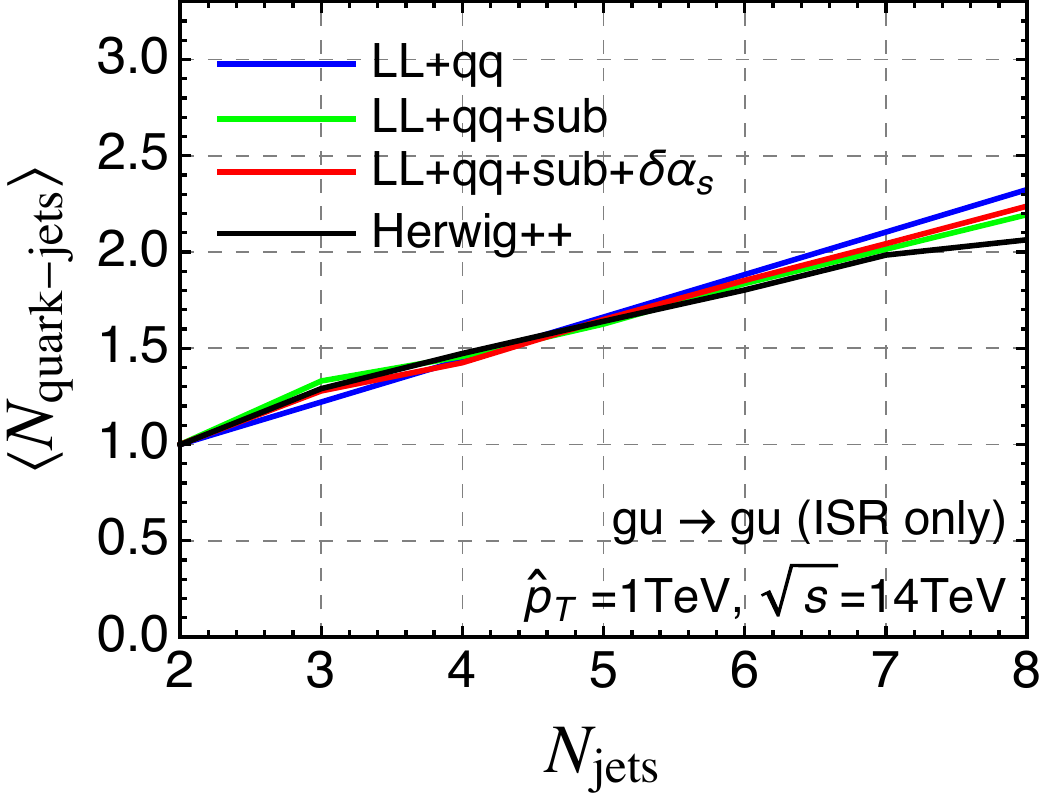}
\includegraphics[width=0.32\linewidth, bb=0 0 300 230]{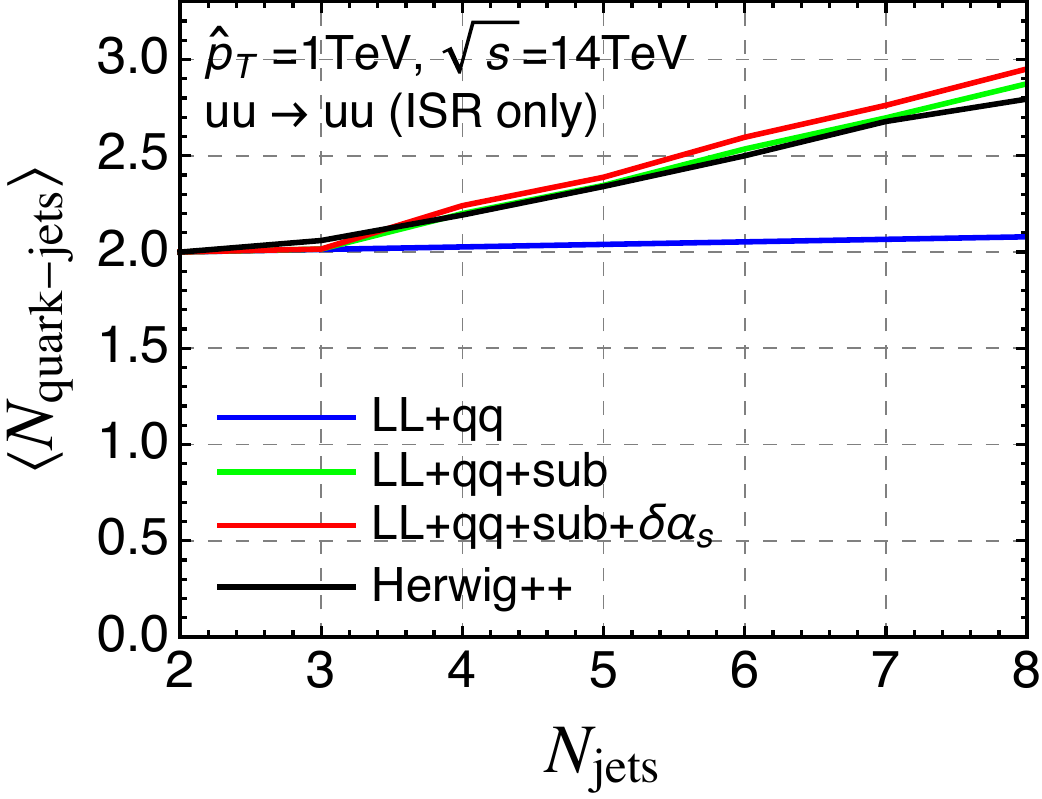}
\caption{{\footnotesize 
Same as Fig.~\ref{fig:Nq_all}, considering only the initial state radiation. 
}}
\label{fig:Nq_isr}
\end{center}
\end{figure}

\begin{figure}[!ht]
\begin{center}
\includegraphics[width=0.32\linewidth, bb=0 0 300 230]{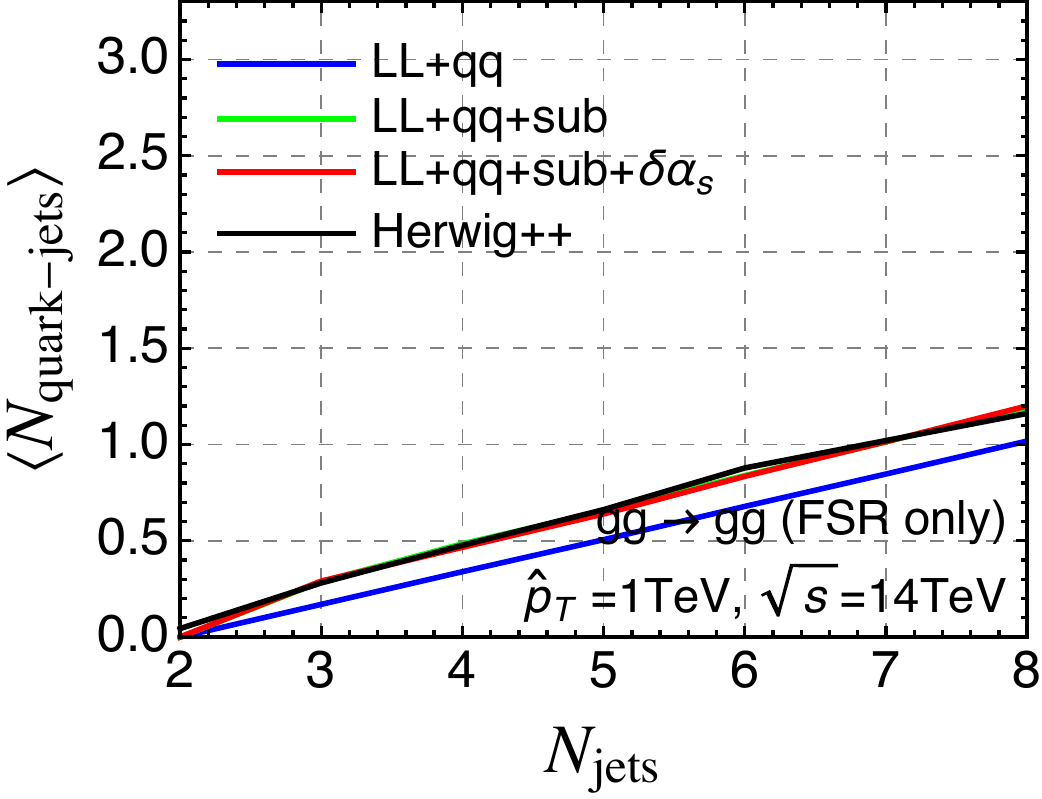}
\includegraphics[width=0.32\linewidth, bb=0 0 300 230]{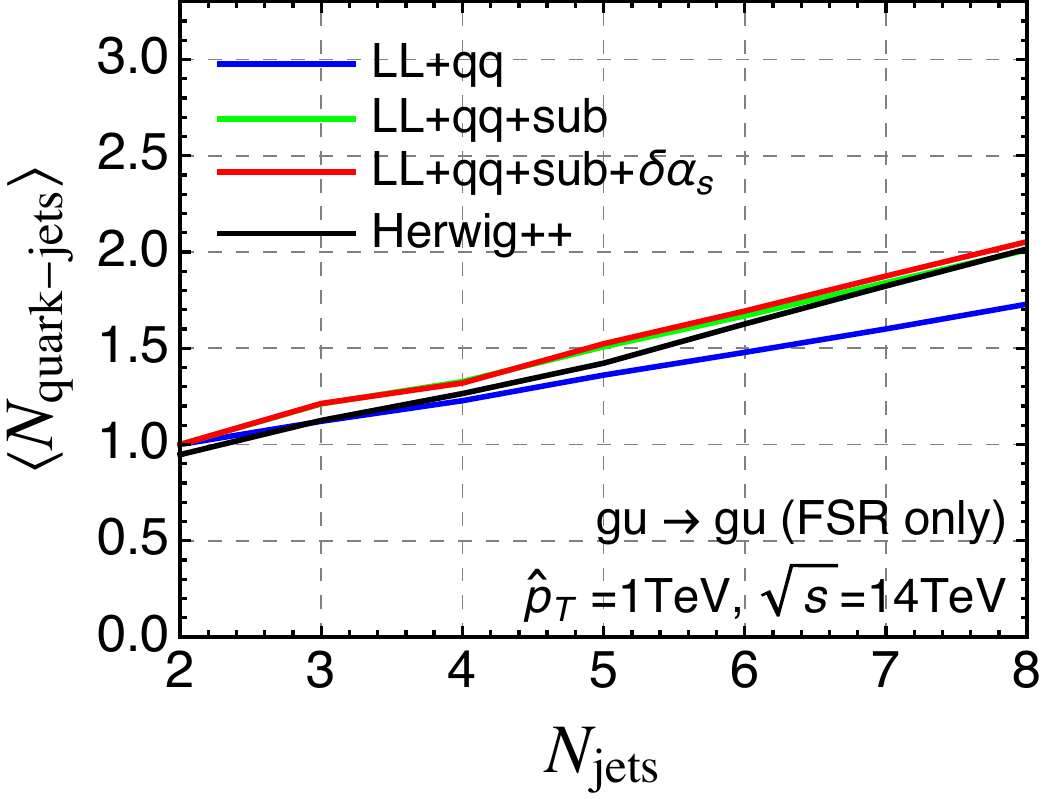}
\includegraphics[width=0.32\linewidth, bb=0 0 300 230]{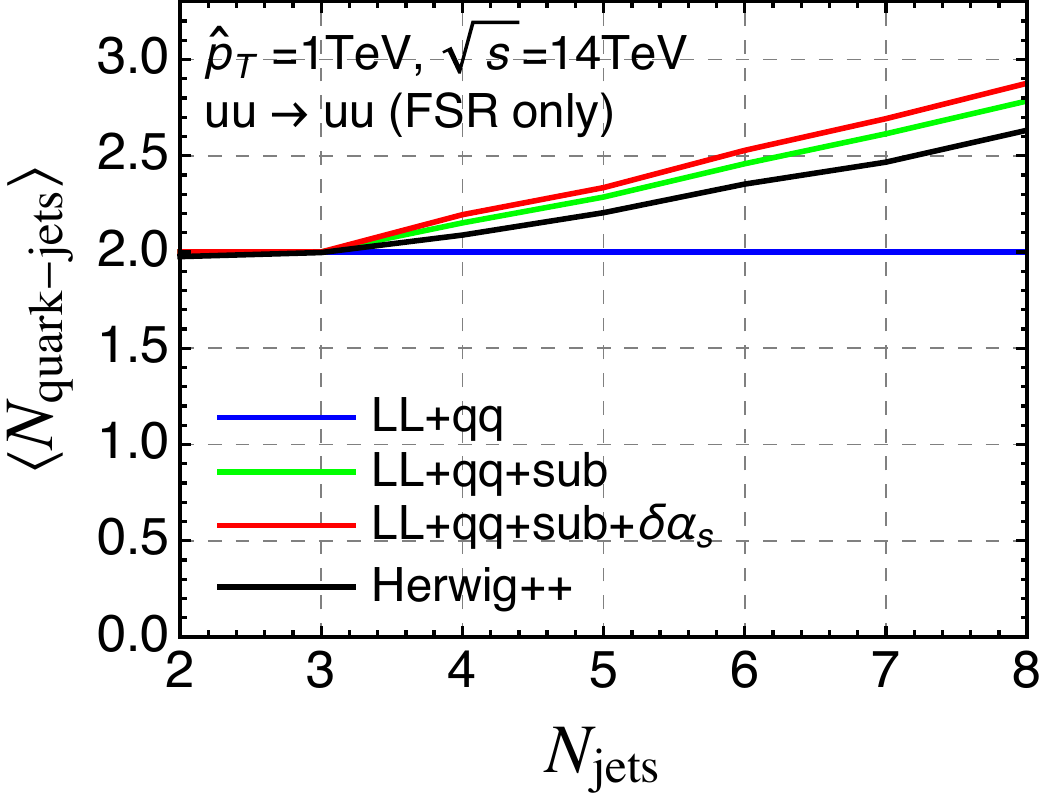}
\caption{{\footnotesize 
Same as Fig.~\ref{fig:Nq_all}, considering only the final state radiation. 
}}
\label{fig:Nq_fsr}
\end{center}
\end{figure}

Generating funtionals for initial state radiations (ISR) and final state radiations (FSR) are calculated in Sec.~\ref{sec:Quark-jet-rates}. 
The number of quark jets are evaluated using the funtionals for three Born configulations in Sec.~\ref{sec:Numerical-results}. 
In the calculation, the contribution from ISR and FSR to the number are combined. 
In Fig.~\ref{fig:Nq_isr} and Fig.~\ref{fig:Nq_fsr}, we show results in which only ISR and FSR are taken into account. 
The results for $gg\to gg$ (left), $gu\to gu$ (center) and $uu\to uu$ (right) are shown. 
In the calculation, $p_0=50$ GeV, and $R=0.4$ are used. 
We set $\hat{p}_T$ to $1$ TeV. 
The blue, green and red curves are the analytical calculations using the functionals labeled by $(\LL+\qq)$, $(\LL+\qq+\sub)$ and $(\LL+\qq+\sub+\delta\alpha_s)$. 
The black curves show the Monte-Carlo prediction given by {\tt Herwig++}.

When we neglect subsequent emissions, the increase of quark jets for $uu\to uu$ in the case of ISR-only is tiny for the same reason as discussed in Fig.~\ref{fig:Nq_all}. 
In the case of FSR-only, the generating functional doesn't contain $v$, so the number of quark jets doesn't increase at all. 
The main cause of the increase of quark jets for $uu\to uu$ stems from $\exp(S_q)$ which is related to subsequent emissions, and the lowest order at which $v$ appears is $\mathcal{O}(u^4v^4)$, therefore, the number of quark jets begins to increase clearly from $\Njets=4$ as discussed in Sec.~\ref{sec:Numerical-results}.  
For the case of ISR-only and $gg\to gg$, the number of quark jets decreases a lot when we take into account the subsequent emissions in our analytic calculation. 
This is mainly because the coefficients for $uv$ in Eq.~(\ref{eq:wbar}) takes a large negative number, which stems from the improvement of approximation to the generating functional ratio in Eq.~(\ref{eq:PsiG1}).


\end{document}